\documentclass{iopjournal}

\usepackage{soul}
\usepackage{float}
\usepackage[caption=false]{subfig} 
\usepackage{gensymb}
\usepackage{graphicx}
\usepackage{dcolumn}
\usepackage{bm}
\usepackage{booktabs}
\usepackage{adjustbox}
\usepackage{booktabs}
\usepackage{longtable}
\usepackage{amssymb}
\usepackage{amsmath}
\usepackage{makecell}
\usepackage{array}
\usepackage{multirow}
\usepackage{natbib}
\usepackage[normalem]{ulem} 

\begin{document}

\articletype{Paper} 

\title{Entropy-based classification of core-collapse supernova gravitational waves for nuclear equation-of-state inference}

\author{
Aknur Sakan$^{1,2}$\orcid{0009-0001-8784-4470}, 
Nurzhan Ussipov$^{1,2,*}$\orcid{0000-0002-2512-3280}, 
Y. Sultan Abylkairov$^{3,4,*}$\orcid{0000-0003-2656-7294},  
Marat Zaidyn$^{1}$\orcid{0009-0006-8505-7277}, 
Alisher Zhunuskanov$^{1}$\orcid{0009-0002-5435-9740},
Almat Akhmetali$^{3,2,1}$\orcid{0009-0005-7254-524X},
José A.~Font$^{5,6}$\orcid{0000-0001-6650-2634}, 
Matthew C. Edwards$^{7}$\orcid{0000-0003-2861-5616}, 
and Ernazar Abdikamalov$^{3,2}$\orcid{0000-0001-5481-7727}
}

\affil{$^1$Department of Electronics and Astrophysics, Al-Farabi Kazakh National University, 050040 Almaty, Kazakhstan}\\
\affil{$^2$Department of Physics, Nazarbayev University, 010000 Astana, Kazakhstan}\\
\affil{$^3$Energetic Cosmos Laboratory, Nazarbayev University, 010000 Astana, Kazakhstan}\\
\affil{$^4$School of Artificial Intelligence and Data Science, Astana IT University, 020000 Astana, Kazakhstan}\\
\affil{$^5$Departamento de Astronomía y Astrofísica, Universitat de València, Avinguda Vicent Andr\'es Estell\'es 19, 46100 Burjassot (Valencia), Spain}\\
\affil{$^6$Observatori Astronòmic, Universitat de València, Catedrático José Beltrán 2, 46980 Paterna, Spain}\\
\affil{$^7$Department of Statistics, University of Auckland, Auckland 1010, New Zealand}\\

\affil{$^{*}$Author to whom any correspondence should be addressed.}

\email{nurzhan.ussipov@kaznu.edu.kz, sultan.abylkairov@nu.edu.kz}


\begin{abstract}

Gravitational waves from core-collapse supernovae provide a unique probe of the physical processes occurring during stellar collapse and of the properties of dense nuclear matter. In this work we investigate the classification of supernova gravitational-wave signals associated with different nuclear equations of state using an entropy-based machine learning framework. The analysis combines signal decomposition, feature extraction, and feature selection techniques to identify informative characteristics of the waveforms and distinguish between different equation of state models. We evaluate the performance of several machine learning classifiers using a compact set of selected features and assess their robustness in the presence of detector noise. We find that entropy-based representations capture the underlying differences between waveform classes and enable accurate classification while maintaining better performance compared to raw data under noisy conditions. These results demonstrate the potential of complexity-based feature extraction as an efficient and interpretable approach for gravitational wave studies of core-collapse supernovae.

\end{abstract}

\section{Introduction}

Einstein’s theory of general relativity predicts the existence of gravitational waves (GWs), subtle distortions that propagate through space-time and are generated by the asymmetric acceleration of massive objects. The first direct detection of GWs, GW150914, was made by Advanced LIGO in 2015 from the merger of two black holes~\citep{abbott2016prl}, opening a new window for observing the universe. Since then, the cumulative total of confirmed GW events detected by the LIGO--Virgo--KAGRA collaboration has reached 390,  
including binary black hole, binary neutron star, and neutron star--black hole mergers~\citep{Abac25GWTC4,LVK-GWTC5}. These observations have established GW astronomy as a powerful tool for probing compact objects and energetic astrophysical phenomena.

Core-collapse supernovae (CCSNe) are another promising source of GWs~\citep{mueller:13, abbott2020ccsne, Kotake17, szczepanczyk2024}. Unlike compact-binary mergers, however, CCSN GWs have not yet been directly detected. The mechanism driving CCSN explosions remains uncertain despite decades of theoretical and numerical study~\citep{muller20hydrodynamics}. GWs could provide a unique probe of this process because they carry information directly from the stellar core, offering insight into the dynamics of the collapse and the formation of the compact remnant~\citep{abdikamalov22GW, mezzacappa24gravitational}.

Entropy, originally introduced in thermodynamics and statistical mechanics, was later adapted by information theory to quantify the degree of uncertainty and predictability in a system \citep{shannon1948}. When applied to signals, entropy-based measures quantify the regularity, complexity, and temporal organization of their patterns, providing a means of distinguishing between different dynamical states. Such measures have been applied extensively to biomedical signals and neurodynamics, among other fields involving complex signals \citep{Kannathal2005, cofre2025entropy, entropy_review}. In the context of GW studies of CCSNe, Summerscales et al. \citep{summerscales2008maximum} demonstrated that maximum entropy reconstruction could, in principle, enable inference of CCSN progenitor properties such as rotation rate and bounce depth for nearby events observed by a network of interferometers. While maximum entropy excels at image-like reconstruction, it struggles to statistically distinguish between the turbulent-driven GW emission and realistic non-Gaussian instrumental noise, otherwise known as glitches \citep{cornish2015bayeswave}. Modern pipelines for data analysis of GW transients (e.g., coherent WaveBurst, cWB; \citep{klimenko2008coherent,klimenko2016method}) and fully Bayesian frameworks (e.g., BayesWave; \citep{cornish2015bayeswave}) superseded it by prioritizing cross-detector coherence and flexible Bayesian model selection, which offer superior robustness in the presence of glitches. 

Recent studies have further explored the use of entropy-based methods to characterize GW signals. For instance, sample and multiscale entropy quantify signal complexity across scales, distinguishing astrophysical signals from noise~\citep{Mohsen21Multiscale}. Conditional entropy and mutual information have been applied for template-free detection of coherent signals across detectors~\citep{ussipov2023information, ussipov2024}, and total correlation has been used to improve Bayesian evidence estimation for compact binary coalescence events~\citep{Zahraoui26}. Maximum entropy spectral analysis provides accurate power spectral density estimation for short time series~\citep{martini2024maximum}. These efforts demonstrated the conceptual relevance of entropy, but the use of entropy as a feature extraction tool for machine learning (ML) classification remains unexplored. 

A key challenge in applying ML methods to the analysis of GW signals is the extraction of informative and discriminative features from the data. Their morphology is governed by complex astrophysical mechanisms and can vary significantly even among events of similar physical origin \cite{maggiore2008gw}. In addition, the typically low signal-to-noise ratio (SNR), often $\lesssim 10$,  further complicates the analysis of time series and hinder the extraction of distinctive characteristics suitable for ML. For reference, simulated GW signals from Galactic CCSNe observed with second-generation detectors such as Advanced LIGO are expected to have signal-to-noise ratios of order $\mathrm{SNR}\sim10$–$30$, depending on the explosion dynamics and distance \citep{Szczepanczyk21Detecting, gossan16observing}. Consequently, the development of feature extraction methods capable of transforming raw signals into compact and interpretable representations while preserving essential physical information is of central importance for improving ML models and enabling physically meaningful inference.

In this study, we introduce a method to classify CCSN GW signals using entropy-based features. We consider a deliberately simplified yet challenging dataset consisting of bounce signals from rapidly rotating progenitors with different nuclear equations of state (EOS). To enhance the informativeness of the features, we first decompose each signal using Variational Mode Decomposition (VMD), which isolates characteristic oscillatory components from noisy non-stationary signals. We then compute fourteen entropy-based descriptors from these components to quantify signal complexity. To determine which descriptors are most informative for classification, we employ feature selection using the minimum redundancy–maximum relevance (mRMR) criterion. After feature selection, the resulting feature set is used to train and evaluate four classifiers: $k$-nearest neighbors (KNN), support vector machines (SVM), random forest (RF), and Naïve Bayes (NB).

This paper is organized as follows. Section~\ref{sec:astro_context} outlines the astrophysical context of CCSNe and their GW emission. Section~\ref{sec:methods} describes the dataset, the computation of entropy features, the feature selection procedures, and the classification algorithms. Section~\ref{sec:results} presents the main results, and Section~\ref{sec:conclusions} summarizes the key findings and conclusions.

\section{Astrophysical Context}
\label{sec:astro_context}

For completeness, this section provides additional background on CCSNe and their GW emission. Readers primarily interested in the data analysis and entropy-based methods may skip it without loss of continuity.

CCSNe mark the end of massive stars (with initial masses in the range from about 8 to 50 solar masses). When such a star exhausts its nuclear fuel, its core can no longer support itself against gravity, and it collapses inward. The collapse stops once matter reaches nuclear densities, creating a rebound that launches a shock wave. This shock quickly loses energy and stalls \cite{Liebendoerfer01}, but if it is revived, it drives a powerful explosion that tears the star apart \cite{burrows93, Janka01Conditions}. A successful explosion leaves a neutron star remnant, while failed or partially successful ones may form black holes, either directly or through fallback accretion~\cite{Kuroda22, Burrows23Black, Powell25noEMCCSN, Eggenberger25}.

During collapse, the newborn proto–neutron star (PNS) emits an enormous number of neutrinos \cite{rampp00, oconnor13Progenitor}, and a small fraction of them are absorbed in the region behind the stalled shock \cite{Buras06b, Bruenn16, Kotake18}. This heats the gas, drives turbulent motion, and helps push the shock outward~\cite{Abdikamalov15, Radice16Neutrino, Vartanyan22collapse}. Additional turbulence can also come from large-scale instabilities such as the standing accretion shock instability~\cite{blondin03stability, foglizzo06neutrino, mueller:12} or from convective plumes originating in the outer stellar layers~\cite{Couch15Three, Mueller17Supernova, Kazeroni20impact, Telman24Convective}. The latter mechanism, known as the neutrino-driven mechanism, is believed to power explosions in slowly or non-rotating stars, likely the dominant class among massive stars~\cite{Heger05Presupernova, popov:12}.

In rapidly rotating stars, the situation is very different. The forming neutron star gains substantial rotational kinetic energy, which magnetic fields can tap to power the explosion. This results in powerful, magnetorotationally driven jet explosions~\cite{burrows:07b, moesta:14b, kuroda:20}. If the jet fails to penetrate the stellar envelope, it can still contribute to the explosion by depositing energy within the star~\cite{Eisenberg22, Pais23choked}. In moderately rotating stars, it is possible that both neutrino heating and magnetic fields contribute to the explosion dynamics~\cite{Takiwaki16Three, Summa18Rotation, abdikamalov2021, Buellet23Effect, Powell24GW}. 

Gravitational waves provide a unique means of probing the interior of CCSNe. Because they originate directly from the stellar core, they carry information about the physical conditions in the star's innermost regions ~\cite{TorresForne19, mezzacappa24gravitational}. The GW emission mostly arises from oscillations of the PNS and typically dominates the high-frequency component ($\sim300$--$2000$ Hz) during the post-bounce accretion phase ~\cite{Murphy09Model, Morozova18}. Emission from non-radial hydrodynamics instabilities can also contribute at frequencies of $\sim100$--$1000$ Hz, typically developing tens to hundreds of milliseconds after bounce~\cite{ott13, Yakunin15GW, kuroda16, Mezzacappa23Core, Vartanyan23Gravitational}. Additionally, lower-frequency ($\lesssim250$ Hz) components may result from anisotropic neutrino emission throughout the post-bounce evolution \cite{mueller:97, Choi24GW}, asymmetric shock propagation during shock expansion and explosion \cite{mueller12Parametrized, radice:19gw}, or jets during jet launching and propagation~\cite{Birnholtz13GW_jet, Gottlieb23Jetted, Soker23GWJJ}. When the progenitor rotates rapidly, centrifugal forces distort the collapsing core, so that the proto neutron star formation is accompanied by a sharp gravitational wave spike at bounce~\cite{ott12correlated, Fuller15SNseismology} with frequencies of $\sim700$ Hz. This is followed by ring-down oscillations at a similar frequency that lasts for $\sim 10$ ms. Rapid rotation can also trigger non-axisymmetric instabilities that produce long-lasting, nearly periodic GW signals frequencies of $\sim100$--$1000$ Hz beginning soon after bounce ~\cite{Scheidegger08, Shibagaki20new}. In rotating models, an additional enhancement of GW emission may arise from the resonance between the fundamental quadrupolar oscillation mode of the PNS and the epicyclic oscillations at the inner-core boundary during the early post-bounce phase, producing emission in the kHz range~\citep{Cusinato:26}.

Although rapidly rotating CCSNe are expected to be relatively rare, potentially accounting for only $\sim 1\%$ of all events \citep{Heger05Presupernova, woosley:06}, they offer important practical advantages for systematic GW studies. In particular, the rotating bounce and early post-bounce dynamics can be modeled with axisymmetric simulations at comparatively modest computational cost \citep{ott:07cqg, Dimmelmeier08, ott12correlated}. This enables the construction of extensive waveform catalogs spanning a broad range of progenitor structures, rotation rates, and EOS models, providing the large and diverse datasets needed to investigate GW-based inference of core properties and the high-density EOS.

Once detected, these signals can provide direct probe of the supernova dynamics \cite{nunes2024deep, pastor24, Villegas25Parameter, Akhmetali2026b, Akhmetali2026c}, and may also reveal the mass, radius, and rotation rate of the PNS \cite{abdikamalov:14, Bizouard21Inference, Bruel23Inference, CasallasLagos23Characterizing}. Comparing GW patterns across different explosion scenarios could help distinguish between neutrino-driven and magnetorotational supernovae~\cite{Logue12Inferring, Powell24Determining}. Constraining the properties of high-density nuclear matter is another key objective~\cite{richers:17, edwards17}, as GW observations can help discriminate among competing nuclear models~\cite{chao22determining, Wolfe23GW, Murphy24Dependence, abylkairov2025, abylkairov25assessing, Akhmetali2026a}.

The rich physics of CCSNe makes them interesting laboratories for studying stellar physics and astrophysics under extreme conditions \cite{pajkos19, pajkos21, mitra23,mitra24}. As current and upcoming GW detectors approach the sensitivity to capture these signals \cite{gossan16observing, Szczepanczyk21Detecting}, developing robust methods to characterize and classify them becomes increasingly important. The next sections focus on such methods that rely on entropy-based representations of GW signals.


\section{Methods}
\label{sec:methods}

Our methodological approach consists of the following steps: signal decomposition, spectral analysis, entropy-based feature extraction, feature selection, and machine learning classification. Each stage in this methodology is discussed in the subsequent sections. First, we provide an overview of the dataset used in this study.

\subsection{Gravitational Wave Data}

GW signals were obtained from the publicly available dataset \cite{richers:17}. Following our previous works~\cite{mitra24,abylkairov2025,Akhmetali2026a} we use the same four EOS, namely \texttt{SFHo} \cite{steiner:13b}, \texttt{LS220} \cite{lseos:91}, \texttt{HSDD2} \cite{hempel:10,hempel:12}, and \texttt{GShenFSU2.1} \cite{gshen:11b}. These four EOS were selected because they are consistent with experimental and neutron-star mass constraints while producing relatively distinct characteristic GW frequencies, making them suitable for investigating EOS-dependent features in the signal \cite{richers:17, mitra24}. We consider a single progenitor model with zero age main sequence mass of $12\,M_\odot$. Impact of rotation is quantified by the ratio $T/|W|$, where $T$ denotes rotational kinetic energy and $W$ denotes gravitational binding energy, evaluated at core bounce ($t = 0$). In this work, we consider $T/|W|$ values ranging from 0.02 to 0.10, where the lower limit is motivated by weak signals with high numerical noise and the upper limit is selected since at higher values extreme rotation inhibits the core from reaching the high densities where differences in EOS are most pronounced. After this filtering, 146 waveforms remain, with 37, 36, 38, and 35 waveforms for the four EOS classes, respectively, resulting in balanced dataset. All waveform data have a sampling rate of 16384 Hz, which matches Advanced LIGO's operational sampling rate. In this work we consider the signal range from $-2$ to $6$ ms relative to the bounce time, since the signal below $-2$ ms is weak and the signal after $6$ ms includes contributions from prompt convection, which is not accurately modeled in the simulations \cite{richers:17}.

To quantify the noise level presented in the signal we use the SNR \cite{flanagan98}. The matched-filter SNR $\rho$ for a detected GW signal $h$ is defined under the assumption of a perfectly matched template,
\begin{equation}
\rho = \sqrt{\int_0^{\infty}\frac{4\hat{h}^*(f)\hat{h}(f)}{S_n(f)}df},
\end{equation}
where $\hat{h}(f)$ is the Fourier transform of the template waveform and $S_n(f)$ is the one-sided noise spectral density. All SNR values are calculated assuming optimal orientation of both the detector and the source, corresponding to a sky location, polarization, and inclination that maximize the detector response. In this work, we use the Advanced LIGO noise spectrum at its design sensitivity in the absence of glitches, specifically in the high-power, zero-detuning configuration~\citep{barsotti2018a+}.

Since the signal amplitude depends on $T/|W|$ and EOS \cite{richers:17, abdikamalov22GW}, and we aim to classify EOS while being robust to variations in $T/|W|$, we control the SNR by rescaling the noise rather than the signal. For each waveform $h$, we first compute the reference SNR with a fixed noise realization, then rescale the noise to achieve the desired target SNR and add it to the simulated GW strain \cite{abylkairov25assessing}:
\begin{equation}
H = h + n \cdot \frac{\text{SNR}_\mathrm{o}}{\text{SNR}_\mathrm{t}},
\label{eq:snr}
\end{equation}
where $\text{SNR}_\mathrm{o}$ is the original SNR calculated from the injected detector noise, $\text{SNR}_\mathrm{t}$ is the target SNR, and $n$ is the detector noise generated using the \texttt{PyCBC} library \cite{pycbc}. We added colored noise directly without additional frequency-dependent weighting because the Advanced LIGO sensitivity curve is relatively flat in the frequency region where the GW signal has most of its power. In Equation (\ref{eq:snr}), we rescaled the noise rather than the signal to preserve the original distribution of signal amplitudes for a given noise level.

\subsection{Variational Mode Decomposition and Spectral Analysis}
\label{sec:VMD}

Variational Mode Decomposition (VMD)~\citep{dragomiretskiy2014vmd} is an adaptive signal processing technique widely used for the analysis of nonlinear and non-stationary time series. The method decomposes an input signal into a predefined number $K$ of intrinsic mode functions (IMFs), each representing a narrow-band oscillatory component characterized by a distinct center frequency. Unlike recursive decomposition methods, VMD formulates the decomposition as a constrained variational optimization problem, simultaneously estimating the IMFs and their center frequencies while minimizing their spectral bandwidths. The optimization is performed iteratively in the frequency domain using a Wiener filtering scheme~\citep{wiener1949}, resulting in a set of spectrally localized modes that capture different frequency structures present in the signal.

\begin{figure}
 \centering
        \includegraphics[width=0.99\textwidth]{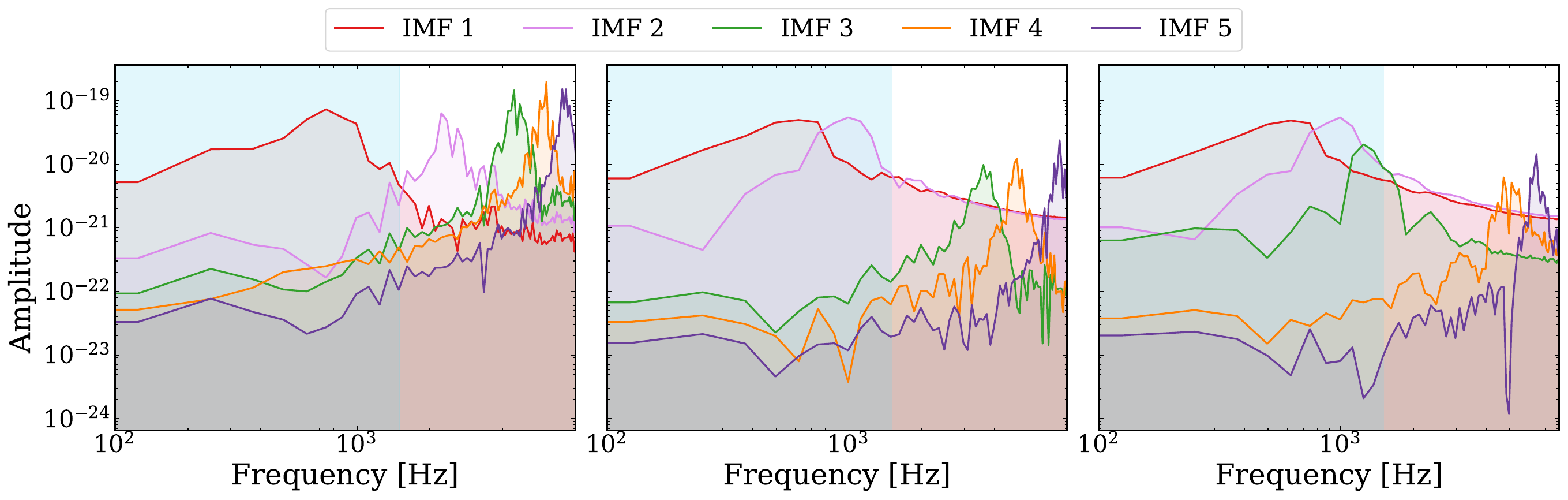}
 \caption{Spectral distribution of the five IMFs for SFHo EOS model with $T/|W|=0.05$ and SNR values of 10, 100, and 200 from left to right panels,  respectively. The shaded light-blue region highlights the frequency window associated with the physical information of signals. IMFs with dominant frequency in shaded region are selected to reconstruct the signal.}
\label{fig_Spec}
\end{figure}

After decomposition, we compute the Fourier transform of each IMF. For every mode, we identify the dominant frequency, $f_{\rm dominant}$, corresponding to the maximum spectral amplitude. We then retain only those IMFs with $f_{\rm dominant}<1500$ Hz, as shown in Figure~\ref{fig_Spec}. This cutoff was chosen to retain the dominant frequency content of the CCSN GW signals, which is concentrated below $\sim1500$ Hz, while excluding higher-frequency components that contribute less to the overall signal. The selected modes are transformed back to the time domain and summed to reconstruct a filtered version of the GW signal. This procedure suppresses high-frequency components that are predominantly noise-dominated while preserving the frequency range that contains the most relevant physical information.

\subsection{Entropy Feature Extraction}
\label{subsec:entropy}

Entropy measures provide a quantitative description of the uncertainty, irregularity, and complexity of a signal. Because CCSN GW signals are highly nonlinear and non-stationary, entropy-based features are well suited for capturing their evolving temporal and spectral characteristics.

Rather than relying on a single entropy measure, we extract a set of 14 entropy features, outlined in Table~\ref{EFtable}, that capture complementary aspects of the signal while limiting redundancy and feature dimensionality. The entropy measures are calculated from the IMF time series obtained through VMD. In the following definitions, \(x\) denotes an individual IMF time series, and the indexing notation (e.g., $i$) refers to the corresponding samples,
histogram bins, decomposition components, or frequency bins, depending
on the entropy measure. Together, these measures characterize temporal organization, statistical complexity, energy distribution, and localized time-frequency structure. This compact feature representation provides a balance between descriptive power and computational efficiency, making it suitable for low-latency classification applications.

For the Rényi ($\alpha$) and Tsallis ($q$) entropy measures, we consider two parameter values, 0.1 and 2. These parameters control the sensitivity of the entropy to the underlying probability distribution. Lower values emphasize low-probability events and increase sensitivity to signal variability, whereas higher values place greater weight on dominant patterns and regular structures. Both measures reduce to Shannon entropy in the limit $\alpha, q \rightarrow 1$.

\begin{longtable}{|p{2.5cm}|p{5.2cm}|p{6.5cm}|}
\caption{Entropy measures used for feature extraction} \label{EFtable} \\

\hline
\textbf{Feature} & \textbf{Formula} & \textbf{Importance} \\ \hline
\endfirsthead

\hline
\textbf{Entropy} & \textbf{Formula} & \textbf{Importance} \\ \hline
\endhead

\hline
\endfoot


\multicolumn{3}{|l|}{\textbf{Information-theoretic measures}} \\ \hline

Shannon Entropy \cite{shannon1948}
& $\mathrm{ShanEn}(x) = -\sum_{i=1} p_i \log_2 p_i$,
\newline
$p_i$ - empirical probability density of the signal amplitude in the $i$-th histogram bin
& Measures uncertainty of the signal based on its probability distribution. Higher entropy indicates more uniform distributions of values. \\ \hline

Rényi Entropy \cite{Renyi1961}
& $\mathrm{RenEn}(x)=\frac{1}{1-\alpha}\log_2\!\left(\sum_{i} p_i^{\alpha}\right)$,
\newline
$\alpha$ - order parameter ($\alpha \neq 1$),
\newline
$p_i$ - empirical probability density of the signal amplitude in the $i$-th histogram bin
& A logarithmic generalization of Shannon with a tunable parameter ($\alpha$). This parameter controls sensitivity to frequent or rare events in the distribution. \\ \hline

Tsallis Entropy \cite{Tsallis1988}
& $\mathrm{TsalEn}(x) = \frac{1-\sum_{i} p_i^{q}}{q-1}$,
\newline
$q$ - order parameter ($q \neq 1$),
\newline
$p_i$ - empirical probability density of the signal amplitude in the $i$-th histogram bin
& A nonlinear (non-additive) generalization of Shannon entropy designed for systems where interactions between components are not independent. It adjusts how probabilities contribute to the total entropy using a non-linear parameter. \\ \hline


\multicolumn{3}{|l|}{\textbf{Energy-based measures}} \\ \hline

Energy Entropy \cite{Yu2006}
& $\mathrm{EnrEn}(x) = -\sum_{i} p_i \log_2 p_i$,
\newline
$p_i = x_i^2/\sum_i x_i^2$ - normalized energy contribution of the $i$-th signal sample
& Describes how signal energy is distributed across components or segments. More uniform distributions correspond to higher entropy. \\ \hline

Logarithmic Energy Entropy \cite{Kara2009}
& $\mathrm{LogEnt}(x) = \sum_i \log(x_i^2)$
& Highlights small amplitude variations by applying a logarithmic transformation to energy values, improving sensitivity to subtle changes. \\ \hline

Dispersion Entropy \cite{Rostaghi2016}
& $\mathrm{DispEn}(x) = -\sum_{i} p_i\log p_i$,
\newline
$p_i$ - probability of the $i$-th dispersion pattern
& Measures complexity by mapping signal values into classes and evaluating the distribution of symbolic patterns, capturing both amplitude and local variations. \\ \hline


\multicolumn{3}{|l|}{\textbf{Transform-based measures}} \\ \hline

Wavelet Entropy \cite{Rosso2001}
& $\mathrm{WavEn}(x) = -\sum_{i} p_i \log_2 p_i$,
\newline
$p_i = E_i/\sum_i E_i$ - normalized energy of the $i$-th wavelet decomposition level
& Quantifies the distribution of energy across wavelet scales, capturing time-frequency variability in non-stationary signals. \\ \hline

Singular Value Decomposition Entropy \cite{SVDEntropy2021}
& $\mathrm{SVDEn}(x) = -\sum_{i=1} \tilde{\sigma}_i \log_2(\tilde{\sigma}_i)$
\newline
$\tilde{\sigma}_i$ - normalization of the singular value obtained from the SVD matrix
& Measures complexity of reconstructed dynamics via the distribution of singular values in phase-space representation. \\ \hline

Spectral Entropy \cite{shannon1948}
& $\mathrm{SpecEn}(x) = -\sum_{i=1} \tilde{f_i(x)} \log (\tilde{f_i(x)})$
\newline
$\tilde{f_i}(x)=f_i(x)/\sum_{i} f_i(x)$ - normalized power spectral density corresponding to the $i$-th frequency componen
& Describes how signal power is distributed across frequency components. Concentrated energy in a few frequencies leads to lower entropy, while distributed energy leads to higher entropy. \\ \hline


\multicolumn{3}{|l|}{\textbf{Embedding-based measures}} \\ \hline

Sample Entropy \cite{Richman2000}
& $\mathrm{SampEn}(x;m,r) = -\log \left(\frac{C(m + 1, r)}{C(m, r)}\right)$,
\newline
$C(m, r)$ - number of matching embedded vectors of length $m$ within tolerance $r$
& Measures the complexity and regularity of a time series by evaluating the conditional probability that similar patterns of length $m$ remain similar when extended to length $m+1$. Higher values indicate more irregular and less predictable signals. \\ \hline

Approximate Entropy \cite{Pincus1991}
& $\mathrm{ApEn}(x;m,r) = \Phi_m(r) - \Phi_{m+1}(r)$,
\newline
$\Phi_m(r)$ - average logarithmic probability that embedded vectors of length $m$
& Measures the regularity of a signal by evaluating how often similar patterns repeat. Lower values indicate more regular and predictable behavior, while higher values indicate more irregular dynamics. \\ \hline

Permutation Entropy \cite{Bandt2002}
& $\mathrm{PermEn}(x;m,\tau) = -\sum_i p_i\log_2 p_i$
\newline
$p_i$ - probability of the $i$-th permutation pattern
& Quantifies complexity based on the ordering of values within short signal segments, focusing on temporal structure rather than amplitude. \\ \hline

\end{longtable}

\subsection{Feature Selection}
\label{subsec:fsandclass}

Feature sets often contain redundant or weakly informative variables that can degrade classification performance and reduce model interpretability. To address this issue, we employ the minimum redundancy--maximum relevance (mRMR) feature selection method \citep{Peng2005}, which ranks features according to their importance and identifies a compact subset of the most discriminative variables.

The mRMR approach is based on mutual information (MI), a measure of statistical dependence between two random variables that quantifies how much information one variable provides about another \citep{kraskov2004}. In the context of classification, MI can be used to assess both the relevance of a feature to the target class and the redundancy between pairs of features. The mRMR criterion seeks features that are highly informative about the class label while containing minimal overlapping information. This objective is expressed as

\begin{equation}
\max \Phi = D - R,
\end{equation}

where $D$ denotes the average mutual information between the selected features and the class labels (maximum relevance), and $R$ represents the average mutual information among the selected features themselves (minimum redundancy).

\subsection{Classification}
\label{sec:classification}

After feature selection, we apply four classical ML algorithms for classification. To ensure a fair comparison between models, we tune each classifier using \texttt{GridSearchCV} with 5-fold cross-validation~\cite{pedregosa2011scikit}. In this procedure, the dataset is divided into five subsets, each consisting of 29 waveforms. For every combination of hyperparameters, the model is trained on four subsets and validated on the remaining one. This process repeats until each fold has served as a validation set and the average classification accuracy across all subsets is computed to evaluate model performance. The combination of hyperparameters that yields the highest average accuracy is selected as optimal. The models obtained during cross-validation are used only to identify the optimal hyperparameters and are not used for subsequent performance evaluation. Once the optimal hyperparameters are determined, each classifier is retrained from scratch using the selected configuration, and only these retrained models are used for the final predictions and reported performance estimates.

\begin{table}[ht]
\centering
\caption{Grid of hyperparameters used for tuning classical ML models. The optimal hyperparameters
are highlighted in bold.}
\label{ML_grid}
\small
\begin{adjustbox}{max width=\textwidth}
\begin{tabular}{p{7cm} | p{7cm}}
\toprule
Random Forest (RF) & Support Vector Machines (SVM) \\
\midrule
`n\_estimators': [50, 75, \textbf{100}, 125, 150] & `C': [0.1, 1, \textbf{10}] \\ 
`max\_depth': [None, 10, 15, \textbf{20}] & `kernel': [`linear', \textbf{`rbf'}, `poly'] \\ 
`min\_samples\_split': [\textbf{2}, 5, 10] & `gamma': [\textbf{`scale'}, `auto'] \\ 
`min\_samples\_leaf': [\textbf{1}, 2, 4] & `degree': [\textbf{2}, 3, 4] \\
\midrule
Na\"{i}ve Bayes (NB) & $k$-Nearest Neighbors (KNN) \\
\midrule
`var\_smoothing': [$10^{-9}$, $10^{-8}$, $10^{-7}$, $\mathbf{10^{-6}}$, $10^{-5}$] & `n\_neighbors': [\textbf{3}, 5, 7, 9, 11] \\
 & `weights': [`uniform', \textbf{`distance'}] \\
 & `metric': [\textbf{`euclidean'}, `manhattan', `minkowski'] \\
 & `p': [\textbf{1}, 2] \\
\bottomrule
\end{tabular}
\end{adjustbox}
\end{table}

To verify that the tuned models generalize well, the mean accuracy on the training folds is compared with that on validation folds. The absence of a significant performance gap indicates that the models are neither overfitting nor underfitting. The resulting optimal hyperparameters for each model, obtained using \texttt{GridSearchCV}, are summarized and highlighted in Table \ref{ML_grid}.

After determining the optimal hyperparameters, the dataset is split into training and test sets with an 80:20 ratio. This split is performed separately from the training and validation sets used during hyperparameter optimization. For each train/test realization, all preprocessing steps and the mRMR feature selection are performed using the training set only, and the resulting transformations and selected features are then applied unchanged to the test set. Model performance is evaluated based on accuracy, which quantifies the proportion of correctly predicted samples relative to the total number of predictions, defined as:

\begin{equation}
\text{Accuracy} = \frac{\text{Number of Correct Predictions}}{\text{Total Number of Predictions}}.
\end{equation}

\begin{figure}[h!]
    \centering
    \includegraphics[width=0.9\textwidth]{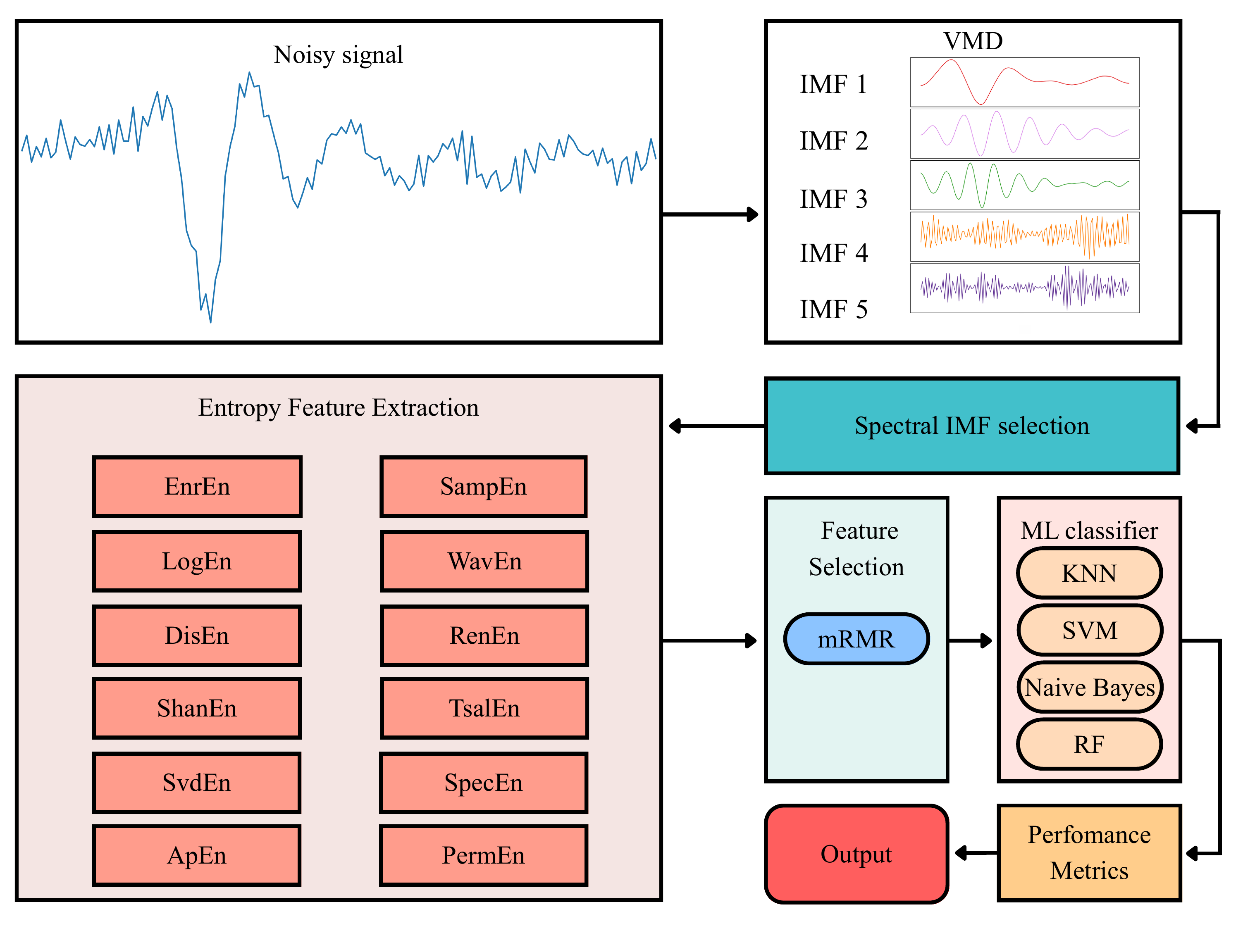}
    \caption{Flowchart outlining the main steps of the proposed methodology. The signal decomposition stage is described in Section \ref{sec:VMD}, the entropy feature extraction in Section \ref{subsec:entropy}, the feature selection method and the classification procedure in Section \ref{subsec:fsandclass}.}
    \label{fig2}
\end{figure}

A summary of the overall methodology is provided in Figure~\ref{fig2}.

\section{Results}
\label{sec:results}

All gravitational wave signals are first preprocessed using VMD and the spectral selection procedure described in Section~\ref{sec:VMD} to suppress noise and retain the components associated with the core dynamics. To quantify these dynamics and assess their sensitivity to the nuclear EOS, we extract a set of entropy-based features from the reconstructed signals.

\begin{figure}
 \centering
        \includegraphics[width=0.92\textwidth]{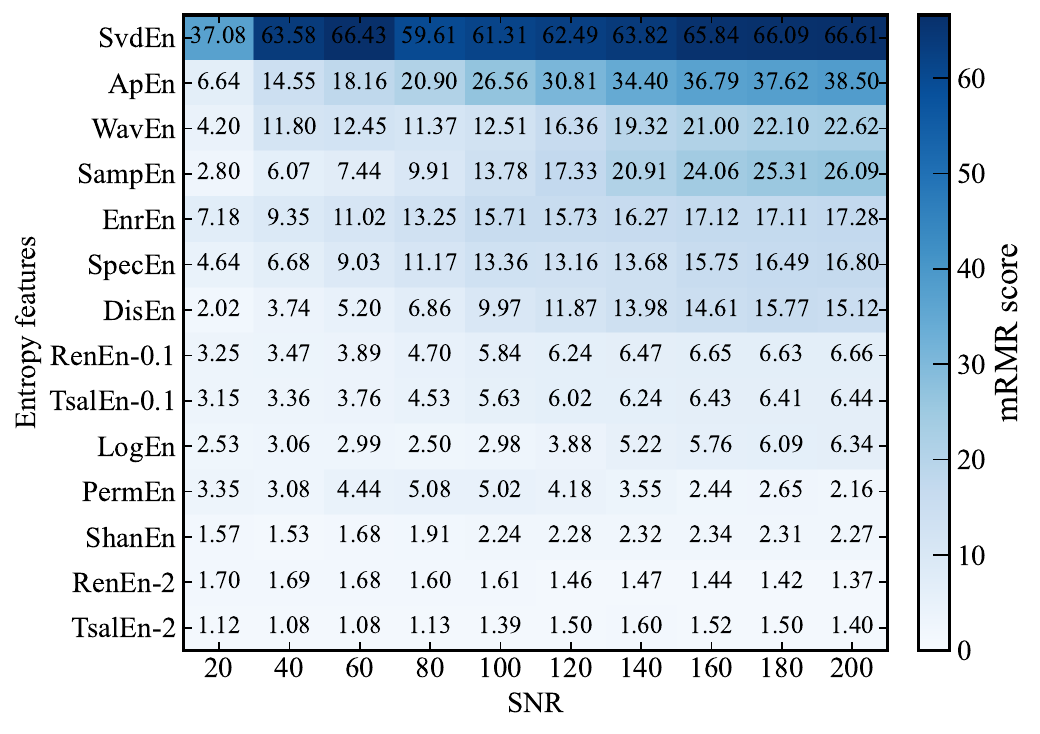}
\caption{Ranking of entropy features based on mRMR scores across varying SNR levels.}
\label{fig_heatmap}
\end{figure}

Figure~\ref{fig_heatmap} presents the fourteen entropy measures ranking obtained by mRMR algorithm across different SNRs ranging from 20 to 200. Due to the working logic of the mRMR algorithm, the absolute value of the score should not be interpreted as a proportional measure of importance. With this caveat, the ranking remains informative about which complementary measures the classifier relies on. Several measures, including SVDEn, ApEn, WavEn, and SampEn, consistently exhibit high ranking across all SNR levels.

All high ranked entropy measures share common characteristic: they probe the temporal regularity, phase-space structure, or time–frequency energy distribution of the signal rather than its amplitude distribution alone. This finding is consistent with previous studies showing that EOS information encoded in the morphology of the waveform, its oscillatory shape and characteristic frequencies, and is therefore best captured by measures sensitive to temporal and spectral structure \citep{abylkairov2025,richers:17,Akhmetali2026a}. 

In contrast, all low ranked entropy probes the amplitude probability distribution and are invariant under reordering of the time samples. They are therefore insensitive to the oscillatory morphology that carries the EOS information, and they are mutually redundant. 

\begin{table}[H]
\centering
\caption{Classification accuracy (mean ± std, \%) for SVM, KNN, RF, and NB over 100 runs at SNR = 50, 100, and 200 using three inputs: time series, full entropy-based features (extracted using $K=3$ IMFs), and mRMR-selected features ($K=3$ IMFs)}
\begin{tabular}{l lcccc}
\hline
SNR & Method & SVM & RF & KNN & NB \\
\hline

\multirow{3}{*}{50} 
& Time series & $55.00 \pm 7.87$ & $43.67 \pm 7.31$ & $36.87 \pm 8.65$ & $35.93 \pm 7.58$ \\
& Entropy-based features & $74.20 \pm 7.45$ & $70.13 \pm 7.21$ & $62.07 \pm 10.6$ & $50.60 \pm 9.51$ \\
& mRMR feature selection & $73.73 \pm 7.82$ & $72.13 \pm 7.91$ & $61.60 \pm 8.75$ & $54.93 \pm 9.34$ \\

\hline

\multirow{3}{*}{100} 
& Time series & $75.70 \pm 6.47$ & $60.87 \pm 7.92$ & $52.80 \pm 7.88$ & $47.80 \pm 8.46$ \\
& Entropy-based features & $78.53 \pm 7.85$ & $72.00 \pm 8.46$ & $65.83 \pm 8.84$ & $52.47 \pm 8.35$ \\
& mRMR feature selection & $79.70 \pm 8.10$ & $72.10 \pm 8.16$ & $67.33 \pm 8.30$ & $53.73 \pm 9.44$ \\

\hline

\multirow{3}{*}{200} 
& Time series & $91.67 \pm 5.63$ & $79.10 \pm 7.39$ & $72.70 \pm 6.71$ & $60.07 \pm 11.3$ \\
& Entropy-based features & $87.97 \pm 5.94$ & $83.33 \pm 7.86$ & $80.30 \pm 6.63$ & $59.17 \pm 9.37$ \\
& mRMR feature selection & $92.23 \pm 4.55$ & $82.67 \pm 6.85$ & $83.97 \pm 6.83$ & $63.13 \pm 7.96$ \\

\hline
\end{tabular}
\label{tab:class}
\end{table}

\begin{figure}
 \centering
        \includegraphics[width=0.85\textwidth]{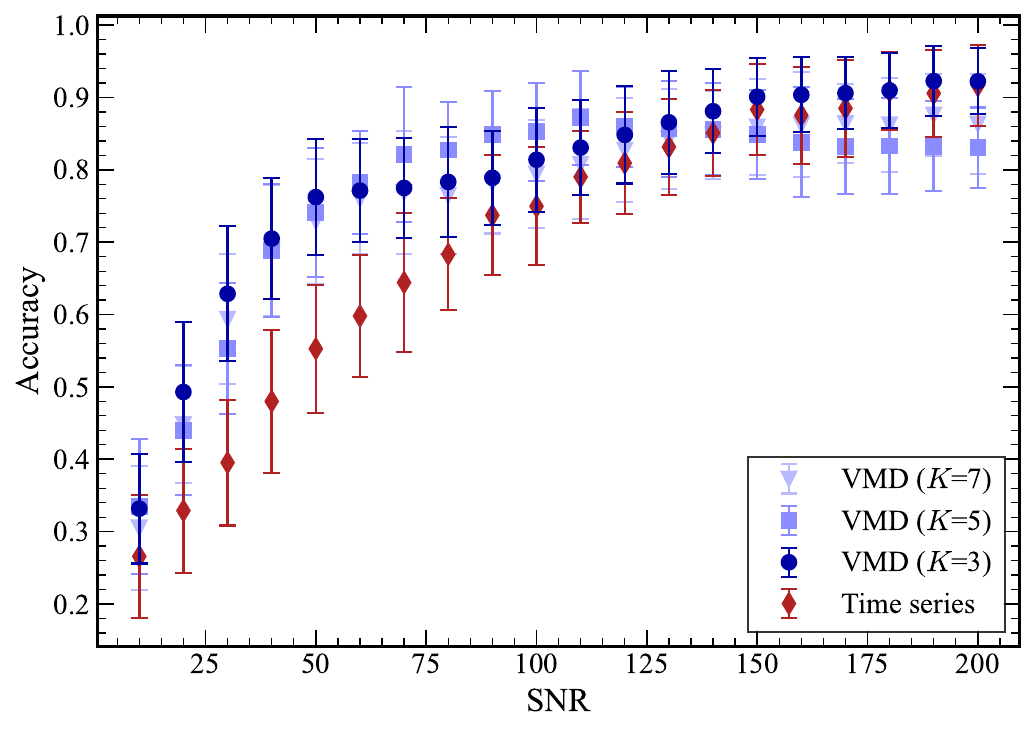}
 \caption{SVM classification accuracy across varying SNR levels. Blue curves represent entropy-based features extracted from VMD with K = 3 (dark blue), K = 5 (medium blue), and K = 7 (light blue). Red curves correspond to classification using the full time-series data.}
\label{fig_ACC}
\end{figure}

Next, we evaluate the classification performance using both the full set of 14 entropy features and a reduced set of 10 features obtained by excluding the lowest-performing measures (PermEn, ShanEn, Renyi-2, and Tsallis-2). We consider four classifiers—SVM, KNN, RF, and NB—and assess their performance at SNR levels of 50, 100, and 200. The results are summarized in Table~\ref{tab:class}. Across all classifiers and SNR values, the full and reduced feature sets yield broadly comparable accuracies, indicating that the excluded features contribute little discriminative information. 
The observed differences are generally within the standard deviations, suggesting no systematic benefit or degradation from excluded entropy features after feature selection.

We further compare these entropy-based representations with a time-series-based baseline using the same set of classifiers. SVM achieves the best overall performance in both cases, reaching 91.67\% accuracy with the time-series input and 92.23\% with the reduced entropy feature set at SNR=200. This trend remains consistent at lower SNR values (50 and 100), where SVM continues to outperform the other classifiers. Overall, entropy-based features generally perform better than raw time-series inputs, particularly at lower SNRs. This may be attributed to the fact that the entropy measures summarize different aspects of the signal's temporal structure into a compact set of features, making the representation less sensitive to individual noisy samples than the raw time series. Based on these results, SVM is selected for all subsequent analyses.

To assess the effect of signal decomposition on classification performance, we evaluate SVM accuracy for different numbers of VMD modes. In VMD, the parameter $K$ controls the number of IMFs into which the signal is decomposed. Figure~\ref{fig_ACC} presents the classification accuracy as a function of SNR for different values of $K$ ($K=3$, 5, and 7), along with a baseline using the raw time-series representation.

In all cases, the accuracy increases with SNR, as expected. However, the choice of $K$ has a noticeable impact on performance. In particular, using three modes ($K=3$) yields the highest or comparable accuracy across most SNR values. Using fewer modes may provide a more compact representation of the dominant signal characteristics, while additional modes can introduce less informative features that do not improve classification. The consistent performance of $K=3$ therefore suggests that three modes are sufficient to capture the relevant information in the signals. We therefore adopt $K=3$ as the default configuration for all evaluations, including the entropy feature representations reported in Table~\ref{tab:class}.

Using this optimal 3-IMF representation, we next compare the SVM performance obtained with entropy-based features against that obtained from the raw time series. As SNR increases, the two approaches converge in performance. At lower SNR values, however, the entropy-based representation provides a clear advantage. For instance, at SNR = 30, which is close to what is expected from a Galactic CCSN, the entropy-based features achieve an accuracy of 62.9\%, compared to 39.5\% for the time-series input. This demonstrates that the entropy-based representation retains useful discriminative information even in the lower-SNR regime relevant to realistic Galactic CCSN observations.

\begin{figure}
 \centering
        \includegraphics[width=0.85\textwidth]{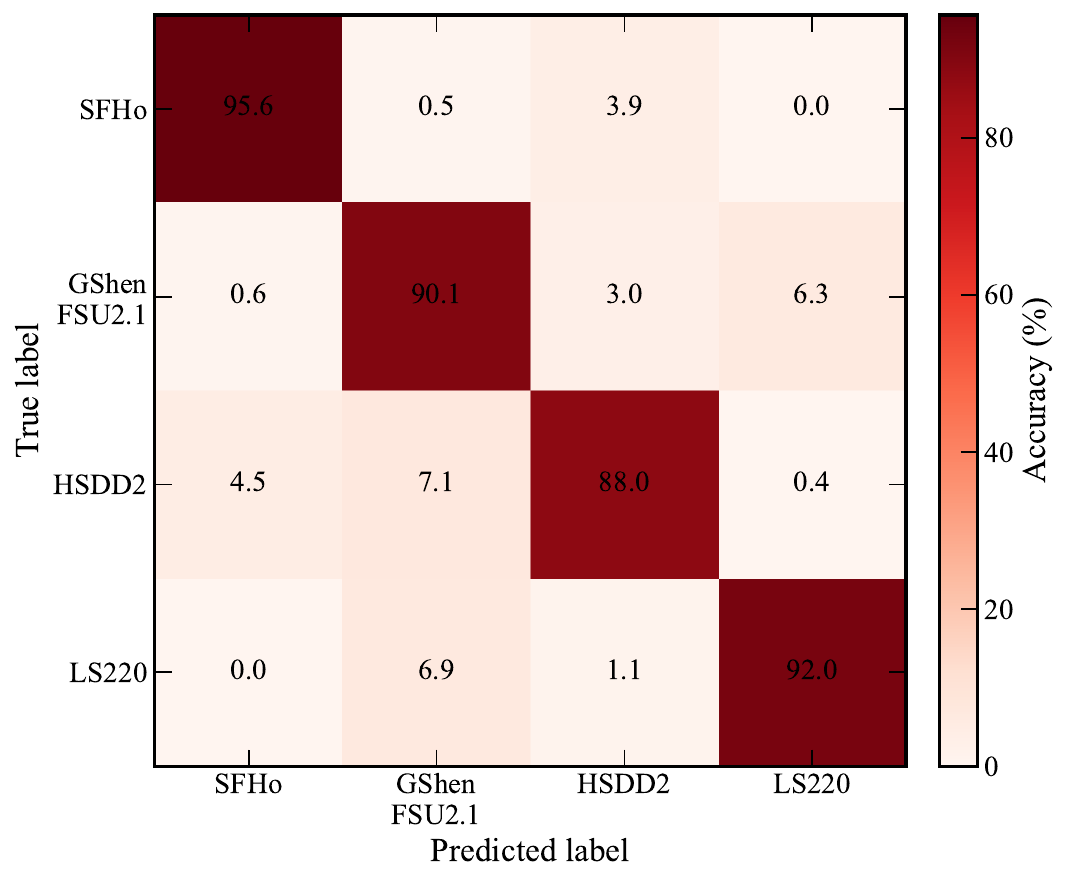}
\caption{Normalized confusion matrix over 100 independent runs for EOS classification at an SNR of 200 using an SVM trained on the selected entropy features. The diagonal elements represent the classification accuracy for each class, while the off-diagonal elements indicate the corresponding misclassification rates. Colorbar represents the classification accuracy.}
\label{fig_CM}
\end{figure}

Finally, we evaluate the class-wise classification performance using the normalized confusion matrix shown in Fig.~\ref{fig_CM}. The model achieves an overall classification accuracy of $91.44\%$, with class-wise accuracies ranging from $88.00\%$ to $95.63\%$. The standard deviation of $3.23\%$ indicates a relatively small variation in performance across the four classes, suggesting that the classifier performs consistently without a strong bias toward any particular class.

Overall, combining VMD-based decomposition with entropy feature extraction yields a compact and noise-robust representation that provides good classification performance across a wide range of SNRs.

\section{Conclusions}
\label{sec:conclusions}

In this study, we have developed an entropy-based machine learning framework for classifying gravitational wave signals from core-collapse supernovae with different nuclear equations of state. The approach combines variational mode decomposition and spectral analysis for signal denoising, followed by entropy-based feature extraction and mRMR feature selection to construct a compact representation of the waveform.

Analysis of feature importance across a wide range of SNRs shows that a reduced subset of entropy measures carries most of the discriminative information. In particular, features related to temporal
regularity, phase-space structure, or time–frequency energy distribution of the signal dominate the classification performance, indicating that the imprint of the equation of state is primarily encoded in the morphology of the waveform.

Using this reduced feature representation, an SVM classifier achieves an accuracy of up to 92.23\% and providing a clear advantage at low SNR compared to raw time series. Overall, the results demonstrate that a small set of well-chosen entropy descriptors is sufficient to capture the relevant physical differences between models, providing an efficient and noise-robust framework for gravitational wave signal classification. However, these classification accuracies should be interpreted within the scope of the simplified waveform set considered in this study and should not be taken as representative of the full diversity of realistic CCSN GW signals.

Several limitations remain to be addressed. Our analysis is restricted to rapidly rotating CCSN models and considers four EOS classes out of the 18 EOSs available in the original dataset, while the diversity of realistic progenitors and nuclear EOSs is substantially broader. In addition, the analysis focuses on the GW signal around core bounce and the early post-bounce phase, and therefore does not capture the full duration and complexity of CCSN GW emission. The detector noise is also assumed to be stationary and Gaussian, and thus does not capture the non-stationary and non-Gaussian noise transients present in real detector data. Future work should address these limitations by considering a broader range of EOSs and progenitor rotation rates, including non-rotating and slowly rotating models, as well as more realistic detector noise containing transient glitches. Extending the analysis to realistic detector data and the full CCSN GW signal will provide a more stringent test of the robustness and applicability of the proposed entropy-based classification framework.

\ack{We thank the anonymous referee for the constructive comments and suggestions that helped improve the clarity and quality of this paper.}

\funding{This research was funded by the Science Committee of the Ministry of Science and Higher Education of the Republic of Kazakhstan (Grant No. AP26103591) and partially supported by the Nazarbayev University Faculty Development Competitive Research Grant Program (no. 040225FD4713). 
JAF is funded by the Spanish Agencia Estatal de Investigación (grant PID2024-159689NB-C21) and by the Generalitat Valenciana (Prometeo excellence programme grant CIPROM/2022/49).}

\data{The data used in this study is publicly available at the following URL/DOI: \url{https://doi.org/10.5281/zenodo.201145}}

\bibliographystyle{unsrt}  
\bibliography{references}  

@article{barsotti2018a+,
  title={The A+ design curve},
  author={Barsotti, Lisa and McCuller, Lee and Evans, Matthew and Fritschel, Peter},
  journal={LIGO Document},
  volume={1800042},
  pages={2018},
  year={2018}
}

@software{pycbc,
  author       = {Alex Nitz and
                  Ian Harry and
                  Duncan Brown and
                  Christopher M. Biwer and
                  Josh Willis and
                  Tito Dal Canton and
                  Collin Capano and
                  Thomas Dent and
                  Larne Pekowsky and
                  Gareth S Cabourn Davies and
                  Soumi De and
                  Miriam Cabero and
                  Shichao Wu and
                  Andrew R. Williamson and
                  Bernd Machenschalk and
                  Duncan Macleod and
                  Francesco Pannarale and
                  Prayush Kumar and
                  Steven Reyes and
                  dfinstad and
                  Sumit Kumar and
                  Márton Tápai and
                  Leo Singer and
                  Praveen Kumar and
                  veronica-villa and
                  maxtrevor and
                  Bhooshan Uday Varsha Gadre and
                  Sebastian Khan and
                  Stephen Fairhurst and
                  Arthur Tolley},
  title        = {gwastro/pycbc: v2.3.3 release of PyCBC},
  month        = jan,
  year         = 2024,
  publisher    = {Zenodo},
  version      = {v2.3.3},
  doi          = {10.5281/zenodo.10473621},
  url          = {https://doi.org/10.5281/zenodo.10473621},
}

@article{flanagan98,
  title = {Measuring gravitational waves from binary black hole coalescences. I. Signal to noise for inspiral, merger, and ringdown},
  author = {Flanagan, \'Eanna \'E. and Hughes, Scott A.},
  journal = {Phys. Rev. D},
  volume = {57},
  issue = {8},
  pages = {4535--4565},
  numpages = {0},
  year = {1998},
  month = {Apr},
  publisher = {American Physical Society},
  doi = {10.1103/PhysRevD.57.4535},
  url = {https://link.aps.org/doi/10.1103/PhysRevD.57.4535}
}

@ARTICLE{abbott2016prl,
       author = {{Abbott}, B.~P. and et al.},
        title = "{Observation of Gravitational Waves from a Binary Black Hole Merger}",
      journal = {\prl},
         year = 2016,
        month = feb,
       volume = {116},
       number = {6},
          eid = {061102},
        pages = {061102},
          doi = {10.1103/PhysRevLett.116.061102},
archivePrefix = {arXiv},
       eprint = {1602.03837},
 primaryClass = {gr-qc},
       adsurl = {https://ui.adsabs.harvard.edu/abs/2016PhRvL.116f1102A}
}

@ARTICLE{Abac25GWTC4,
       author = {{The LIGO Scientific Collaboration} and {the Virgo Collaboration} and {the KAGRA Collaboration} and {Abac}, A.~G. and {Abouelfettouh}, I. and {Acernese}, F. and {Ackley}, K. et al},
        title = "{GWTC-4.0: Updating the Gravitational-wave Transient Catalog with Observations from the First Part of the Fourth LIGO--Virgo--KAGRA Observing Run}",
      journal = {\apjl},
         year = 2026,
        month = jun,
       volume = {1004},
       number = {2},
          eid = {L22},
        pages = {L22},
          doi = {10.3847/2041-8213/ae2c74},
       adsurl = {https://ui.adsabs.harvard.edu/abs/2026ApJ..1004L..22A}
}

@ARTICLE{Cusinato:26,
       author = {{Cusinato}, Marco and {Obergaulinger}, Martin and {Aloy}, Miguel {\'A}. and {Font}, Jos{\'e} A.},
        title = "{Resonant amplification of multimessenger emission in rotating stellar core collapse}",
      journal = {Physical Review Research},
         year = 2026,
        month = feb,
       volume = {8},
       number = {1},
          eid = {013180},
        pages = {013180},
          doi = {10.1103/6wp3-sw9y},
       adsurl = {https://ui.adsabs.harvard.edu/abs/2026PhRvR...8a3180C}
}

@ARTICLE{LVK-GWTC5,
       author = {{The LIGO Scientific Collaboration} and {the Virgo Collaboration} and {the KAGRA Collaboration} and {Abac}, A.~G. and et al},
        title = "{GWTC-5.0: Observations from the Second Part of the Fourth LIGO-Virgo-KAGRA Observing Run and Updates to the Gravitational-Wave Transient Catalog}",
      journal = {arXiv e-prints},
         year = 2026,
        month = may,
          eid = {arXiv:2605.27225},
        pages = {arXiv:2605.27225},
          doi = {10.48550/arXiv.2605.27225},
archivePrefix = {arXiv},
       eprint = {2605.27225},
 primaryClass = {gr-qc},
       adsurl = {https://ui.adsabs.harvard.edu/abs/2026arXiv260527225T}
}

@ARTICLE{mueller:13,
       author = {{M{\"u}ller}, Bernhard and {Janka}, Hans-Thomas and {Marek}, Andreas},
        title = "{A New Multi-dimensional General Relativistic Neutrino Hydrodynamics Code of Core-collapse Supernovae. III. Gravitational Wave Signals from Supernova Explosion Models}",
      journal = {\apj},
         year = 2013,
        month = mar,
       volume = {766},
       number = {1},
          eid = {43},
        pages = {43},
          doi = {10.1088/0004-637X/766/1/43},
archivePrefix = {arXiv},
       eprint = {1210.6984},
 primaryClass = {astro-ph.SR},
       adsurl = {https://ui.adsabs.harvard.edu/abs/2013ApJ...766...43M}
}

@article{abbott2020ccsne,
       author = {{Abbott}, B.~P. and et al.},
        title = "{Optically targeted search for gravitational waves emitted by core-collapse supernovae during the first and second observing runs of advanced LIGO and advanced Virgo}",
      journal = {\prd},
         year = 2020,
        month = apr,
       volume = {101},
       number = {8},
          eid = {084002},
        pages = {084002},
          doi = {10.1103/PhysRevD.101.084002},
archivePrefix = {arXiv},
       eprint = {1908.03584},
 primaryClass = {astro-ph.HE},
       adsurl = {https://ui.adsabs.harvard.edu/abs/2020PhRvD.101h4002A}
}

@INCOLLECTION{Kotake17,
       author = {{Kotake}, Kei and {Kuroda}, Takami},
        title = "{Gravitational Waves from Core-Collapse Supernovae}",
    booktitle = {Handbook of Supernovae},
         year = 2017,
       editor = {{Alsabti}, Athem W. and {Murdin}, Paul},
        pages = {1671},
          doi = {10.1007/978-3-319-21846-5_9},
       adsurl = {https://ui.adsabs.harvard.edu/abs/2017hsn..book.1671K}
}

@ARTICLE{muller20hydrodynamics,
       author = {{M{\"u}ller}, Bernhard},
        title = "{Hydrodynamics of core-collapse supernovae and their progenitors}",
      journal = {Living Reviews in Computational Astrophysics},
         year = 2020,
        month = jun,
       volume = {6},
       number = {1},
          eid = {3},
        pages = {3},
          doi = {10.1007/s41115-020-0008-5},
archivePrefix = {arXiv},
       eprint = {2006.05083},
 primaryClass = {astro-ph.SR},
       adsurl = {https://ui.adsabs.harvard.edu/abs/2020LRCA....6....3M}
}

@INCOLLECTION{abdikamalov22GW,
       author = {{Abdikamalov}, Ernazar and {Pagliaroli}, Giulia and {Radice}, David},
        title = "{Gravitational Waves from Core-Collapse Supernovae}",
    booktitle = {Handbook of Gravitational Wave Astronomy},
         year = 2022,
       editor = {{Bambi}, Cosimo and {Katsanevas}, Stavros and {Kokkotas}, Konstantinos D.},
          eid = {21},
        pages = {21},
          doi = {10.1007/978-981-15-4702-7_21-1},
       adsurl = {https://ui.adsabs.harvard.edu/abs/2022hgwa.bookE..21A}
}

@ARTICLE{mezzacappa24gravitational,
       author = {{Mezzacappa}, Anthony and {Zanolin}, Michele},
        title = "{Gravitational Waves from Neutrino-Driven Core Collapse Supernovae: Predictions, Detection, and Parameter Estimation}",
      journal = {arXiv e-prints},
         year = 2024,
        month = jan,
          eid = {arXiv:2401.11635},
        pages = {arXiv:2401.11635},
          doi = {10.48550/arXiv.2401.11635},
archivePrefix = {arXiv},
       eprint = {2401.11635},
 primaryClass = {astro-ph.HE},
       adsurl = {https://ui.adsabs.harvard.edu/abs/2024arXiv240111635M}
}

@ARTICLE{szczepanczyk2024,
       author = {{Szczepa{\'n}czyk}, Marek J. and et al.},
        title = "{Optically targeted search for gravitational waves emitted by core-collapse supernovae during the third observing run of Advanced LIGO and Advanced Virgo}",
      journal = {\prd},
         year = 2024,
        month = aug,
       volume = {110},
       number = {4},
          eid = {042007},
        pages = {042007},
          doi = {10.1103/PhysRevD.110.042007},
archivePrefix = {arXiv},
       eprint = {2305.16146},
 primaryClass = {astro-ph.HE},
       adsurl = {https://ui.adsabs.harvard.edu/abs/2024PhRvD.110d2007S}
}

@article{cornish2015bayeswave,
       author = {{Cornish}, Neil J. and {Littenberg}, Tyson B.},
        title = "{Bayeswave: Bayesian inference for gravitational wave bursts and instrument glitches}",
      journal = {Classical and Quantum Gravity},
         year = 2015,
        month = jul,
       volume = {32},
       number = {13},
          eid = {135012},
        pages = {135012},
          doi = {10.1088/0264-9381/32/13/135012},
archivePrefix = {arXiv},
       eprint = {1410.3835},
 primaryClass = {gr-qc},
       adsurl = {https://ui.adsabs.harvard.edu/abs/2015CQGra..32m5012C}
}

@article{klimenko2016method,
  title={Method for detection and reconstruction of gravitational wave transients with networks of advanced detectors},
  author={Klimenko, S. et al.},
  journal={Physical Review D},
  volume={93},
  number={4},
  pages={042004},
  year={2016}
}

@article{Mohsen21Multiscale,
    author = {Javaherian, Mohsen and Mollaei, Saeid},
    title = {Multiscale Entropy Analysis of Gravitational Waves},
    journal = {Advances in High Energy Physics},
    volume = {2021},
    number = {1},
    pages = {6643546},
    doi = {https://doi.org/10.1155/2021/6643546},
    url = {https://onlinelibrary.wiley.com/doi/abs/10.1155/2021/6643546},
    eprint = {https://onlinelibrary.wiley.com/doi/pdf/10.1155/2021/6643546},
    year = {2021}
}

@article{ussipov2023information,
    title={INFORMATION-ENTROPY METHOD FOR DETECTING GRAVITATIONAL WAVE SIGNALS}, volume={20},
    url={https://phtj.buketov.edu.kz/EPTJ/article/view/721}, 
    DOI={10.31489/2023No2/79-86}, 
    number={2(44)}, 
    journal={Eurasian Physical Technical Journal}, 
    author={Zhanabaev, Z.Zh. and Ussipov, N.M.}, 
    year={2023}, 
    month={Jul.}, 
    pages={79–86} 
}

@ARTICLE{ussipov2024,
       author = {{Ussipov}, Nurzhan and {Zhanabaev}, Zeinulla and {Akhmetali}, Almat and {Zaidyn}, Marat and {Turlykozhayeva}, Dana and {Akniyazova}, Aigerim and {Namazbayev}, Timur},
        title = "{Classification of Gravitational Waves from Black Hole-Neutron Star Mergers with Machine Learning}",
      journal = {Journal of Astronomy and Space Sciences},
         year = 2024,
        month = sep,
       volume = {41},
       number = {3},
        pages = {149-158},
          doi = {10.5140/JASS.2024.41.3.149},
       adsurl = {https://ui.adsabs.harvard.edu/abs/2024JASS...41..149U}
}

@article{martini2024maximum,
  author  = {Martini, A. and Schmidt, S. and Ashton, G. and Del Pozzo, W.},
  title   = {Maximum Entropy Spectral Analysis: an application to gravitational-wave data analysis},
  journal = {Eur. Phys. J. C},
  year    = {2024},
  volume  = {84},
  number  = {10},
  pages   = {1023},
  doi     = {10.1140/epjc/s10052-024-10323-x}
}

@ARTICLE{ott12correlated,
       author = {{Ott}, C.~D. and {Abdikamalov}, E. and {O'Connor}, E. and
         {Reisswig}, C. and {Haas}, R. and {Kalmus}, P. and {Drasco}, S. and
         {Burrows}, A. and {Schnetter}, E.},
        title = "{Correlated gravitational wave and neutrino signals from general-relativistic rapidly rotating iron core collapse}",
      journal = {\prd},
         year = 2012,
        month = jul,
       volume = {86},
       number = {2},
          eid = {024026},
        pages = {024026},
          doi = {10.1103/PhysRevD.86.024026},
archivePrefix = {arXiv},
       eprint = {1204.0512},
 primaryClass = {astro-ph.HE},
       adsurl = {https://ui.adsabs.harvard.edu/abs/2012PhRvD..86b4026O}
}

@ARTICLE{gossan16observing,
       author = {{Gossan}, S.~E. and {Sutton}, P. and {Stuver}, A. and {Zanolin}, M. and
         {Gill}, K. and {Ott}, C.~D.},
        title = "{Observing gravitational waves from core-collapse supernovae in the advanced detector era}",
      journal = {\prd},
         year = 2016,
        month = feb,
       volume = {93},
       number = {4},
          eid = {042002},
        pages = {042002},
          doi = {10.1103/PhysRevD.93.042002},
archivePrefix = {arXiv},
       eprint = {1511.02836},
 primaryClass = {astro-ph.HE},
       adsurl = {https://ui.adsabs.harvard.edu/abs/2016PhRvD..93d2002G}
}

@ARTICLE{Vartanyan22collapse,
       author = {{Vartanyan}, David and {Coleman}, Matthew S.~B. and {Burrows}, Adam},
        title = "{The collapse and three-dimensional explosion of three-dimensional massive-star supernova progenitor models}",
      journal = {\mnras},
         year = 2022,
        month = mar,
       volume = {510},
       number = {4},
        pages = {4689-4705},
          doi = {10.1093/mnras/stab3702},
archivePrefix = {arXiv},
       eprint = {2109.10920},
 primaryClass = {astro-ph.SR},
       adsurl = {https://ui.adsabs.harvard.edu/abs/2022MNRAS.510.4689V}
}

@ARTICLE{Powell24Determining,
       author = {{Powell}, Jade and {Iess}, Alberto and {Llorens-Monteagudo}, Miquel and {Obergaulinger}, Martin and {M{\"u}ller}, Bernhard and {Torres-Forn{\'e}}, Alejandro and {Cuoco}, Elena and {Font}, Jos{\'e} A.},
        title = "{Determining the core-collapse supernova explosion mechanism with current and future gravitational-wave observatories}",
      journal = {\prd},
         year = 2024,
        month = mar,
       volume = {109},
       number = {6},
          eid = {063019},
        pages = {063019},
          doi = {10.1103/PhysRevD.109.063019},
archivePrefix = {arXiv},
       eprint = {2311.18221},
 primaryClass = {astro-ph.HE},
       adsurl = {https://ui.adsabs.harvard.edu/abs/2024PhRvD.109f3019P}
}

@ARTICLE{kuroda:20,
       author = {{Kuroda}, Takami and {Arcones}, Almudena and {Takiwaki}, Tomoya and {Kotake}, Kei},
        title = "{Magnetorotational Explosion of a Massive Star Supported by Neutrino Heating in General Relativistic Three-dimensional Simulations}",
      journal = {\apj},
         year = 2020,
        month = jun,
       volume = {896},
       number = {2},
          eid = {102},
        pages = {102},
          doi = {10.3847/1538-4357/ab9308},
archivePrefix = {arXiv},
       eprint = {2003.02004},
 primaryClass = {astro-ph.HE},
       adsurl = {https://ui.adsabs.harvard.edu/abs/2020ApJ...896..102K}
}

@ARTICLE{Murphy09Model,
       author = {{Murphy}, Jeremiah W. and {Ott}, Christian D. and {Burrows}, Adam},
        title = "{A Model for Gravitational Wave Emission from Neutrino-Driven Core-Collapse Supernovae}",
      journal = {\apj},
         year = 2009,
        month = dec,
       volume = {707},
       number = {2},
        pages = {1173-1190},
          doi = {10.1088/0004-637X/707/2/1173},
archivePrefix = {arXiv},
       eprint = {0907.4762},
 primaryClass = {astro-ph.SR},
       adsurl = {https://ui.adsabs.harvard.edu/abs/2009ApJ...707.1173M}
}

@ARTICLE{popov:12,
       author = {{Popov}, S.~B. and {Turolla}, R.},
        title = "{Initial spin periods of neutron stars in supernova remnants}",
      journal = {\apss},
         year = 2012,
        month = oct,
       volume = {341},
       number = {2},
        pages = {457-464},
          doi = {10.1007/s10509-012-1100-z},
archivePrefix = {arXiv},
       eprint = {1204.0632},
 primaryClass = {astro-ph.HE},
       adsurl = {https://ui.adsabs.harvard.edu/abs/2012Ap&SS.341..457P}
}

@ARTICLE{TorresForne19,
       author = {{Torres-Forn{\'e}}, Alejandro and {Cerd{\'a}-Dur{\'a}n}, Pablo and {Obergaulinger}, Martin and {M{\"u}ller}, Bernhard and {Font}, Jos{\'e} A.},
        title = "{Universal Relations for Gravitational-Wave Asteroseismology of Protoneutron Stars}",
      journal = {\prl},
         year = 2019,
        month = aug,
       volume = {123},
       number = {5},
          eid = {051102},
        pages = {051102},
          doi = {10.1103/PhysRevLett.123.051102},
archivePrefix = {arXiv},
       eprint = {1902.10048},
 primaryClass = {gr-qc},
       adsurl = {https://ui.adsabs.harvard.edu/abs/2019PhRvL.123e1102T}
}

@ARTICLE{Powell24GW,
       author = {{Powell}, Jade and {M{\"u}ller}, Bernhard},
        title = "{The gravitational-wave emission from the explosion of a 15 solar mass star with rotation and magnetic fields}",
      journal = {MNRAS},
         year = 2024,
        month = aug,
       volume = {532},
       number = {4},
        pages = {4326-4339},
          doi = {10.1093/mnras/stae1731},
archivePrefix = {arXiv},
       eprint = {2406.09691},
 primaryClass = {astro-ph.HE},
       adsurl = {https://ui.adsabs.harvard.edu/abs/2024MNRAS.532.4326P}
}

@ARTICLE{Villegas25Parameter,
       author = {{Villegas}, L.~O. and {Moreno}, C. and {Pajkos}, M.~A. and {Zanolin}, M. and {Antelis}, J.~M.},
        title = "{Parameter estimation from the core-bounce phase of rotating core collapse supernovae in real interferometer noise}",
      journal = {Classical and Quantum Gravity},
         year = 2025,
        month = jun,
       volume = {42},
       number = {11},
          eid = {115001},
        pages = {115001},
          doi = {10.1088/1361-6382/add235},
       adsurl = {https://ui.adsabs.harvard.edu/abs/2025CQGra..42k5001V}
}

@ARTICLE{Bizouard21Inference,
       author = {{Bizouard}, Marie-Anne and {Maturana-Russel}, Patricio and {Torres-Forn{\'e}}, Alejandro and {Obergaulinger}, Martin and {Cerd{\'a}-Dur{\'a}n}, Pablo and {Christensen}, Nelson and {Font}, Jos{\'e} A. and {Meyer}, Renate},
        title = "{Inference of protoneutron star properties from gravitational-wave data in core-collapse supernovae}",
      journal = {\prd},
         year = 2021,
        month = mar,
       volume = {103},
       number = {6},
          eid = {063006},
        pages = {063006},
          doi = {10.1103/PhysRevD.103.063006},
archivePrefix = {arXiv},
       eprint = {2012.00846},
 primaryClass = {gr-qc},
       adsurl = {https://ui.adsabs.harvard.edu/abs/2021PhRvD.103f3006B}
}

@ARTICLE{Bruel23Inference,
       author = {{Bruel}, Tristan and {Bizouard}, Marie-Anne and {Obergaulinger}, Martin and {Maturana-Russel}, Patricio and {Torres-Forn{\'e}}, Alejandro and {Cerd{\'a}-Dur{\'a}n}, Pablo and {Christensen}, Nelson and {Font}, Jos{\'e} A. and {Meyer}, Renate},
        title = "{Inference of protoneutron star properties in core-collapse supernovae from a gravitational-wave detector network}",
      journal = {\prd},
         year = 2023,
        month = apr,
       volume = {107},
       number = {8},
          eid = {083029},
        pages = {083029},
          doi = {10.1103/PhysRevD.107.083029},
archivePrefix = {arXiv},
       eprint = {2301.10019},
 primaryClass = {astro-ph.HE},
       adsurl = {https://ui.adsabs.harvard.edu/abs/2023PhRvD.107h3029B}
}

@ARTICLE{Wolfe23GW,
       author = {{Wolfe}, Noah E. and {Fr{\"o}hlich}, Carla and {Miller}, Jonah M. and {Torres-Forn{\'e}}, Alejandro and {Cerd{\'a}-Dur{\'a}n}, Pablo},
        title = "{Gravitational Wave Eigenfrequencies from Neutrino-driven Core-collapse Supernovae}",
      journal = {\apj},
         year = 2023,
        month = sep,
       volume = {954},
       number = {2},
          eid = {161},
        pages = {161},
          doi = {10.3847/1538-4357/ace693},
archivePrefix = {arXiv},
       eprint = {2303.16962},
 primaryClass = {astro-ph.HE},
       adsurl = {https://ui.adsabs.harvard.edu/abs/2023ApJ...954..161W}
}

@ARTICLE{edwards17,
       author = {{Edwards}, Matthew C.},
        title = "{Classifying the equation of state from rotating core collapse gravitational waves with deep learning}",
      journal = {\prd},
         year = 2021,
        month = jan,
       volume = {103},
       number = {2},
          eid = {024025},
        pages = {024025},
          doi = {10.1103/PhysRevD.103.024025},
archivePrefix = {arXiv},
       eprint = {2009.07367},
 primaryClass = {astro-ph.IM},
       adsurl = {https://ui.adsabs.harvard.edu/abs/2021PhRvD.103b4025E}
}

@ARTICLE{nunes2024deep,
       author = {{Nunes}, Solange and {Escrig}, Gabriel and {Freitas}, Osvaldo G. and {Font}, Jos{\'e} A. and {Fernandes}, Tiago and {Onofre}, Antonio and {Torres-Forn{\'e}}, Alejandro},
        title = "{Deep-learning classification and parameter inference of rotational core-collapse supernovae}",
      journal = {\prd},
         year = 2024,
        month = sep,
       volume = {110},
       number = {6},
          eid = {064037},
        pages = {064037},
          doi = {10.1103/PhysRevD.110.064037},
archivePrefix = {arXiv},
       eprint = {2403.04938},
 primaryClass = {astro-ph.HE},
       adsurl = {https://ui.adsabs.harvard.edu/abs/2024PhRvD.110f4037N}
}

@ARTICLE{Logue12Inferring,
       author = {{Logue}, J. and {Ott}, C.~D. and {Heng}, I.~S. and {Kalmus}, P. and {Scargill}, J.~H.~C.},
        title = "{Inferring core-collapse supernova physics with gravitational waves}",
      journal = {\prd},
         year = 2012,
        month = aug,
       volume = {86},
       number = {4},
          eid = {044023},
        pages = {044023},
          doi = {10.1103/PhysRevD.86.044023},
archivePrefix = {arXiv},
       eprint = {1202.3256},
 primaryClass = {gr-qc},
       adsurl = {https://ui.adsabs.harvard.edu/abs/2012PhRvD..86d4023L}
}

@ARTICLE{Janka01Conditions,
       author = {{Janka}, H.-Th.},
        title = "{Conditions for shock revival by neutrino heating in core-collapse supernovae}",
      journal = {\aap},
         year = 2001,
        month = mar,
       volume = {368},
        pages = {527-560},
          doi = {10.1051/0004-6361:20010012},
archivePrefix = {arXiv},
       eprint = {astro-ph/0008432},
 primaryClass = {astro-ph},
       adsurl = {https://ui.adsabs.harvard.edu/abs/2001A&A...368..527J}
}

@ARTICLE{Buras06b,
       author = {{Buras}, R. and {Janka}, H.-Th. and {Rampp}, M. and {Kifonidis}, K.},
        title = "{Two-dimensional hydrodynamic core-collapse supernova simulations with spectral neutrino transport. II. Models for different progenitor stars}",
      journal = {\aap},
         year = 2006,
        month = oct,
       volume = {457},
       number = {1},
        pages = {281-308},
          doi = {10.1051/0004-6361:20054654},
archivePrefix = {arXiv},
       eprint = {astro-ph/0512189},
 primaryClass = {astro-ph},
       adsurl = {https://ui.adsabs.harvard.edu/abs/2006A&A...457..281B}
}

@ARTICLE{Bruenn16,
       author = {{Bruenn}, Stephen W. and {Lentz}, Eric J. and {Hix}, W. Raphael and {Mezzacappa}, Anthony and {Harris}, J. Austin and {Messer}, O.~E. Bronson and {Endeve}, Eirik and {Blondin}, John M. and {Chertkow}, Merek Austin and {Lingerfelt}, Eric J. and {Marronetti}, Pedro and {Yakunin}, Konstantin N.},
        title = "{The Development of Explosions in Axisymmetric Ab Initio Core-collapse Supernova Simulations of 12-25 M Stars}",
      journal = {\apj},
         year = 2016,
        month = feb,
       volume = {818},
       number = {2},
          eid = {123},
        pages = {123},
          doi = {10.3847/0004-637X/818/2/123},
archivePrefix = {arXiv},
       eprint = {1409.5779},
 primaryClass = {astro-ph.SR},
       adsurl = {https://ui.adsabs.harvard.edu/abs/2016ApJ...818..123B}
}

@ARTICLE{Eggenberger25,
       author = {{Eggenberger Andersen}, Oliver and {O'Connor}, Evan and {Andresen}, Haakon and {da Silva Schneider}, Andr{\'e} and {Couch}, Sean M.},
        title = "{Black Hole Supernovae, Their Equation of State Dependence, and Ejecta Composition}",
      journal = {\apj},
         year = 2025,
        month = feb,
       volume = {980},
       number = {1},
          eid = {53},
        pages = {53},
          doi = {10.3847/1538-4357/ada899},
archivePrefix = {arXiv},
       eprint = {2411.11969},
 primaryClass = {astro-ph.HE},
       adsurl = {https://ui.adsabs.harvard.edu/abs/2025ApJ...980...53E}
}

@ARTICLE{rampp00,
       author = {{Rampp}, Markus and {Janka}, H.-Thomas},
        title = "{Spherically Symmetric Simulation with Boltzmann Neutrino Transport of Core Collapse and Postbounce Evolution of a 15 M$_{solar}$ Star}",
      journal = {\apjl},
         year = 2000,
        month = aug,
       volume = {539},
       number = {1},
        pages = {L33-L36},
          doi = {10.1086/312837},
archivePrefix = {arXiv},
       eprint = {astro-ph/0005438},
 primaryClass = {astro-ph},
       adsurl = {https://ui.adsabs.harvard.edu/abs/2000ApJ...539L..33R}
}

@ARTICLE{Abdikamalov15,
       author = {{Abdikamalov}, Ernazar and {Ott}, Christian D. and {Radice}, David and {Roberts}, Luke F. and {Haas}, Roland and {Reisswig}, Christian and {M{\"o}sta}, Philipp and {Klion}, Hannah and {Schnetter}, Erik},
        title = "{Neutrino-driven Turbulent Convection and Standing Accretion Shock Instability in Three-dimensional Core-collapse Supernovae}",
      journal = {\apj},
         year = 2015,
        month = jul,
       volume = {808},
       number = {1},
          eid = {70},
        pages = {70},
          doi = {10.1088/0004-637X/808/1/70},
archivePrefix = {arXiv},
       eprint = {1409.7078},
 primaryClass = {astro-ph.HE},
       adsurl = {https://ui.adsabs.harvard.edu/abs/2015ApJ...808...70A}
}

@ARTICLE{Radice16Neutrino,
       author = {{Radice}, David and {Ott}, Christian D. and {Abdikamalov}, Ernazar and {Couch}, Sean M. and {Haas}, Roland and {Schnetter}, Erik},
        title = "{Neutrino-driven Convection in Core-collapse Supernovae: High-resolution Simulations}",
      journal = {\apj},
         year = 2016,
        month = mar,
       volume = {820},
       number = {1},
          eid = {76},
        pages = {76},
          doi = {10.3847/0004-637X/820/1/76},
archivePrefix = {arXiv},
       eprint = {1510.05022},
 primaryClass = {astro-ph.HE},
       adsurl = {https://ui.adsabs.harvard.edu/abs/2016ApJ...820...76R}
}

@ARTICLE{mueller:12,
       author = {{M{\"u}ller}, Bernhard and {Janka}, Hans-Thomas and {Heger}, Alexander},
        title = "{New Two-dimensional Models of Supernova Explosions by the Neutrino-heating Mechanism: Evidence for Different Instability Regimes in Collapsing Stellar Cores}",
      journal = {\apj},
         year = 2012,
        month = dec,
       volume = {761},
       number = {1},
          eid = {72},
        pages = {72},
          doi = {10.1088/0004-637X/761/1/72},
archivePrefix = {arXiv},
       eprint = {1205.7078},
 primaryClass = {astro-ph.SR},
       adsurl = {https://ui.adsabs.harvard.edu/abs/2012ApJ...761...72M}
}

@ARTICLE{oconnor13Progenitor,
       author = {{O'Connor}, Evan and {Ott}, Christian D.},
        title = "{The Progenitor Dependence of the Pre-explosion Neutrino Emission in Core-collapse Supernovae}",
      journal = {\apj},
         year = 2013,
        month = jan,
       volume = {762},
       number = {2},
          eid = {126},
        pages = {126},
          doi = {10.1088/0004-637X/762/2/126},
archivePrefix = {arXiv},
       eprint = {1207.1100},
 primaryClass = {astro-ph.HE},
       adsurl = {https://ui.adsabs.harvard.edu/abs/2013ApJ...762..126O}
}

@ARTICLE{Burrows93,
       author = {{Burrows}, Adam and {Goshy}, John},
        title = "{A Theory of Supernova Explosions}",
      journal = {\apjl},
         year = 1993,
        month = oct,
       volume = {416},
        pages = {L75},
          doi = {10.1086/187074},
       adsurl = {https://ui.adsabs.harvard.edu/abs/1993ApJ...416L..75B}
}

@ARTICLE{abdikamalov:14,
       author = {{Abdikamalov}, Ernazar and {Gossan}, Sarah and {DeMaio}, Alexandra M. and
         {Ott}, Christian D.},
        title = "{Measuring the angular momentum distribution in core-collapse supernova progenitors with gravitational waves}",
      journal = {\prd},
         year = 2014,
        month = aug,
       volume = {90},
       number = {4},
          eid = {044001},
        pages = {044001},
          doi = {10.1103/PhysRevD.90.044001},
archivePrefix = {arXiv},
       eprint = {1311.3678},
 primaryClass = {astro-ph.SR},
       adsurl = {https://ui.adsabs.harvard.edu/abs/2014PhRvD..90d4001A}
}

@ARTICLE{Kazeroni20impact,
       author = {{Kazeroni}, R{\'e}mi and {Abdikamalov}, Ernazar},
        title = "{The impact of progenitor asymmetries on the neutrino-driven convection in core-collapse supernovae}",
      journal = {\mnras},
         year = 2020,
        month = jun,
       volume = {494},
       number = {4},
        pages = {5360-5373},
          doi = {10.1093/mnras/staa944},
archivePrefix = {arXiv},
       eprint = {1911.08819},
 primaryClass = {astro-ph.SR},
       adsurl = {https://ui.adsabs.harvard.edu/abs/2020MNRAS.494.5360K}
}

@ARTICLE{Mezzacappa23Core,
       author = {{Mezzacappa}, Anthony and {Marronetti}, Pedro and {Landfield}, Ryan E. and {Lentz}, Eric J. and {Murphy}, R. Daniel and {Raphael Hix}, W. and {Harris}, J. Austin and {Bruenn}, Stephen W. and {Blondin}, John M. and {Bronson Messer}, O.~E. and {Casanova}, Jordi and {Kronzer}, Luke L.},
        title = "{Core collapse supernova gravitational wave emission for progenitors of 9.6, 15, and 25M{\ensuremath{\odot}}}",
      journal = {\prd},
         year = 2023,
        month = feb,
       volume = {107},
       number = {4},
          eid = {043008},
        pages = {043008},
          doi = {10.1103/PhysRevD.107.043008},
archivePrefix = {arXiv},
       eprint = {2208.10643},
 primaryClass = {astro-ph.SR},
       adsurl = {https://ui.adsabs.harvard.edu/abs/2023PhRvD.107d3008M}
}

@ARTICLE{Birnholtz13GW_jet,
       author = {{Birnholtz}, Ofek and {Piran}, Tsvi},
        title = "{Gravitational wave memory from gamma ray bursts' jets}",
      journal = {\prd},
         year = 2013,
        month = jun,
       volume = {87},
       number = {12},
          eid = {123007},
        pages = {123007},
          doi = {10.1103/PhysRevD.87.123007},
archivePrefix = {arXiv},
       eprint = {1302.5713},
 primaryClass = {astro-ph.HE},
       adsurl = {https://ui.adsabs.harvard.edu/abs/2013PhRvD..87l3007B}
}

@ARTICLE{Soker23GWJJ,
       author = {{Soker}, Noam},
        title = "{Predicting Gravitational Waves from Jittering-jets-driven Core Collapse Supernovae}",
      journal = {Research in Astronomy and Astrophysics},
         year = 2023,
        month = dec,
       volume = {23},
       number = {12},
          eid = {121001},
        pages = {121001},
          doi = {10.1088/1674-4527/ad013e},
archivePrefix = {arXiv},
       eprint = {2308.04329},
 primaryClass = {astro-ph.HE},
       adsurl = {https://ui.adsabs.harvard.edu/abs/2023RAA....23l1001S}
}

@ARTICLE{Shibagaki20new,
       author = {{Shibagaki}, Shota and {Kuroda}, Takami and {Kotake}, Kei and {Takiwaki}, Tomoya},
        title = "{A new gravitational-wave signature of low-T/|W| instability in rapidly rotating stellar core collapse}",
      journal = {\mnras},
         year = 2020,
        month = mar,
       volume = {493},
       number = {1},
        pages = {L138-L142},
          doi = {10.1093/mnrasl/slaa021},
archivePrefix = {arXiv},
       eprint = {1909.09730},
 primaryClass = {astro-ph.HE},
       adsurl = {https://ui.adsabs.harvard.edu/abs/2020MNRAS.493L.138S}
}

@ARTICLE{Fuller15SNseismology,
       author = {{Fuller}, Jim and {Klion}, Hannah and {Abdikamalov}, Ernazar and {Ott}, Christian D.},
        title = "{Supernova seismology: gravitational wave signatures of rapidly rotating core collapse}",
      journal = {\mnras},
         year = 2015,
        month = jun,
       volume = {450},
       number = {1},
        pages = {414-427},
          doi = {10.1093/mnras/stv698},
archivePrefix = {arXiv},
       eprint = {1501.06951},
 primaryClass = {astro-ph.HE},
       adsurl = {https://ui.adsabs.harvard.edu/abs/2015MNRAS.450..414F}
}

@ARTICLE{Vartanyan23Gravitational,
       author = {{Vartanyan}, David and {Burrows}, Adam and {Wang}, Tianshu and {Coleman}, Matthew S.~B. and {White}, Christopher J.},
        title = "{Gravitational-wave signature of core-collapse supernovae}",
      journal = {\prd},
         year = 2023,
        month = may,
       volume = {107},
       number = {10},
          eid = {103015},
        pages = {103015},
          doi = {10.1103/PhysRevD.107.103015},
archivePrefix = {arXiv},
       eprint = {2302.07092},
 primaryClass = {astro-ph.HE},
       adsurl = {https://ui.adsabs.harvard.edu/abs/2023PhRvD.107j3015V}
}

@ARTICLE{Scheidegger08,
       author = {{Scheidegger}, S. and {Fischer}, T. and {Whitehouse}, S.~C. and {Liebend{\"o}rfer}, M.},
        title = "{Gravitational waves from 3D MHD core collapse simulations}",
      journal = {\aap},
         year = 2008,
        month = oct,
       volume = {490},
       number = {1},
        pages = {231-241},
          doi = {10.1051/0004-6361:20078577},
archivePrefix = {arXiv},
       eprint = {0709.0168},
 primaryClass = {astro-ph},
       adsurl = {https://ui.adsabs.harvard.edu/abs/2008A&A...490..231S}
}

@ARTICLE{moesta:14b,
   author = {{M{\"o}sta}, P. and {Richers}, S. and {Ott}, C.~D. and {Haas}, R. and
	{Piro}, A.~L. and {Boydstun}, K. and {Abdikamalov}, E. and {Reisswig}, C. and
	{Schnetter}, E.},
    title = "{Magnetorotational Core-collapse Supernovae in Three Dimensions}",
  journal = {\apjl},
archivePrefix = "arXiv",
   eprint = {1403.1230},
 primaryClass = "astro-ph.HE",
     year = 2014,
    month = apr,
   volume = 785,
      eid = {L29},
    pages = {L29},
      doi = {10.1088/2041-8205/785/2/L29},
   adsurl = {http://adsabs.harvard.edu/abs/2014ApJ...785L..29M}
}

@ARTICLE{Heger05Presupernova,
       author = {{Heger}, A. and {Woosley}, S.~E. and {Spruit}, H.~C.},
        title = "{Presupernova Evolution of Differentially Rotating Massive Stars Including Magnetic Fields}",
      journal = {\apj},
         year = 2005,
        month = jun,
       volume = {626},
       number = {1},
        pages = {350-363},
          doi = {10.1086/429868},
archivePrefix = {arXiv},
       eprint = {astro-ph/0409422},
 primaryClass = {astro-ph},
       adsurl = {https://ui.adsabs.harvard.edu/abs/2005ApJ...626..350H}
}

@ARTICLE{Pais23choked,
       author = {{Pais}, Matteo and {Piran}, Tsvi and {Nakar}, Ehud},
        title = "{The velocity distribution of outflows driven by choked jets in stellar envelopes}",
      journal = {\mnras},
         year = 2023,
        month = feb,
       volume = {519},
       number = {2},
        pages = {1941-1954},
          doi = {10.1093/mnras/stac3640},
archivePrefix = {arXiv},
       eprint = {2208.14459},
 primaryClass = {astro-ph.HE},
       adsurl = {https://ui.adsabs.harvard.edu/abs/2023MNRAS.519.1941P}
}

@ARTICLE{Buellet23Effect,
       author = {{Buellet}, A. -C. and {Foglizzo}, T. and {Guilet}, J. and {Abdikamalov}, E.},
        title = "{Effect of stellar rotation on the development of post-shock instabilities during core-collapse supernovae}",
      journal = {\aap},
         year = 2023,
        month = jun,
       volume = {674},
          eid = {A205},
        pages = {A205},
          doi = {10.1051/0004-6361/202245799},
archivePrefix = {arXiv},
       eprint = {2301.01962},
 primaryClass = {astro-ph.HE},
       adsurl = {https://ui.adsabs.harvard.edu/abs/2023A&A...674A.205B}
}

@ARTICLE{Summa18Rotation,
       author = {{Summa}, Alexander and {Janka}, Hans-Thomas and {Melson}, Tobias and {Marek}, Andreas},
        title = "{Rotation-supported Neutrino-driven Supernova Explosions in Three Dimensions and the Critical Luminosity Condition}",
      journal = {\apj},
         year = 2018,
        month = jan,
       volume = {852},
       number = {1},
          eid = {28},
        pages = {28},
          doi = {10.3847/1538-4357/aa9ce8},
archivePrefix = {arXiv},
       eprint = {1708.04154},
 primaryClass = {astro-ph.HE},
       adsurl = {https://ui.adsabs.harvard.edu/abs/2018ApJ...852...28S}
}

@ARTICLE{mueller:97,
       author = {{Mueller}, E. and {Janka}, H. -T.},
        title = "{Gravitational radiation from convective instabilities in Type II supernova explosions.}",
      journal = {\aap},
         year = 1997,
        month = jan,
       volume = {317},
        pages = {140-163},
       adsurl = {https://ui.adsabs.harvard.edu/abs/1997A&A...317..140M}
}

@ARTICLE{Choi24GW,
       author = {{Choi}, Lyla and {Burrows}, Adam and {Vartanyan}, David},
        title = "{Gravitational-wave and Gravitational-wave Memory Signatures of Core-collapse Supernovae}",
      journal = {\apj},
         year = 2024,
        month = nov,
       volume = {975},
       number = {1},
          eid = {12},
        pages = {12},
          doi = {10.3847/1538-4357/ad74f8},
archivePrefix = {arXiv},
       eprint = {2408.01525},
 primaryClass = {astro-ph.HE},
       adsurl = {https://ui.adsabs.harvard.edu/abs/2024ApJ...975...12C}
}

@ARTICLE{Powell25noEMCCSN,
       author = {{Powell}, Jade and {M{\"u}ller}, Bernhard},
        title = "{Gravitational waves from core-collapse supernovae with no electromagnetic counterparts}",
      journal = {arXiv e-prints},
         year = 2025,
        month = jun,
          eid = {arXiv:2506.03581},
        pages = {arXiv:2506.03581},
          doi = {10.48550/arXiv.2506.03581},
archivePrefix = {arXiv},
       eprint = {2506.03581},
 primaryClass = {astro-ph.HE},
       adsurl = {https://ui.adsabs.harvard.edu/abs/2025arXiv250603581P}
}

@ARTICLE{pastor24,
       author = {{Pastor-Marcos}, Carlos and {Cerd{\'a}-Dur{\'a}n}, Pablo and {Walker}, Daniel and {Torres-Forn{\'e}}, Alejandro and {Abdikamalov}, Ernazar and {Richers}, Sherwood and {Font}, Jos{\'e} A.},
        title = "{Bayesian inference from gravitational waves in fast-rotating, core-collapse supernovae}",
      journal = {\prd},
         year = 2024,
        month = mar,
       volume = {109},
       number = {6},
          eid = {063028},
        pages = {063028},
          doi = {10.1103/PhysRevD.109.063028},
archivePrefix = {arXiv},
       eprint = {2308.03456},
 primaryClass = {astro-ph.HE},
       adsurl = {https://ui.adsabs.harvard.edu/abs/2024PhRvD.109f3028P}
}

@ARTICLE{Takiwaki16Three,
       author = {{Takiwaki}, Tomoya and {Kotake}, Kei and {Suwa}, Yudai},
        title = "{Three-dimensional simulations of rapidly rotating core-collapse supernovae: finding a neutrino-powered explosion aided by non-axisymmetric flows}",
      journal = {\mnras},
         year = 2016,
        month = sep,
       volume = {461},
       number = {1},
        pages = {L112-L116},
          doi = {10.1093/mnrasl/slw105},
archivePrefix = {arXiv},
       eprint = {1602.06759},
 primaryClass = {astro-ph.HE},
       adsurl = {https://ui.adsabs.harvard.edu/abs/2016MNRAS.461L.112T}
}

@ARTICLE{ott13,
       author = {{Ott}, Christian D. and {Abdikamalov}, Ernazar and {M{\"o}sta}, Philipp and {Haas}, Roland and {Drasco}, Steve and {O'Connor}, Evan P. and {Reisswig}, Christian and {Meakin}, Casey A. and {Schnetter}, Erik},
        title = "{General-relativistic Simulations of Three-dimensional Core-collapse Supernovae}",
      journal = {\apj},
         year = 2013,
        month = may,
       volume = {768},
       number = {2},
          eid = {115},
        pages = {115},
          doi = {10.1088/0004-637X/768/2/115},
archivePrefix = {arXiv},
       eprint = {1210.6674},
 primaryClass = {astro-ph.HE},
       adsurl = {https://ui.adsabs.harvard.edu/abs/2013ApJ...768..115O}
}

@ARTICLE{Szczepanczyk21Detecting,
       author = {{Szczepa{\'n}czyk}, Marek J. and {Antelis}, Javier M. and {Benjamin}, Michael and {Cavagli{\`a}}, Marco and {Gondek-Rosi{\'n}ska}, Dorota and {Hansen}, Travis and {Klimenko}, Sergey and {Morales}, Manuel D. and {Moreno}, Claudia and {Mukherjee}, Soma and {Nurbek}, Gaukhar and {Powell}, Jade and {Singh}, Neha and {Sitmukhambetov}, Satzhan and {Szewczyk}, Pawe{\l} and {Valdez}, Oscar and {Vedovato}, Gabriele and {Westhouse}, Jonathan and {Zanolin}, Michele and {Zheng}, Yanyan},
        title = "{Detecting and reconstructing gravitational waves from the next galactic core-collapse supernova in the advanced detector era}",
      journal = {\prd},
         year = 2021,
        month = nov,
       volume = {104},
       number = {10},
          eid = {102002},
        pages = {102002},
          doi = {10.1103/PhysRevD.104.102002},
archivePrefix = {arXiv},
       eprint = {2104.06462},
 primaryClass = {astro-ph.HE},
       adsurl = {https://ui.adsabs.harvard.edu/abs/2021PhRvD.104j2002S}
}

@article{summerscales2008maximum,
       author = {{Summerscales}, T.~Z. and {Burrows}, Adam and {Finn}, Lee Samuel and {Ott}, Christian D.},
        title = "{Maximum Entropy for Gravitational Wave Data Analysis: Inferring the Physical Parameters of Core-Collapse Supernovae}",
      journal = {\apj},
         year = 2008,
        month = may,
       volume = {678},
       number = {2},
        pages = {1142-1157},
          doi = {10.1086/528362},
archivePrefix = {arXiv},
       eprint = {0704.2157},
 primaryClass = {astro-ph},
       adsurl = {https://ui.adsabs.harvard.edu/abs/2008ApJ...678.1142S}
}

@ARTICLE{mitra23,
       author = {{Mitra}, A. and {Shukirgaliyev}, B. and {Abylkairov}, Y.~S. and {Abdikamalov}, E.},
        title = "{Exploring supernova gravitational waves with machine learning}",
      journal = {\mnras},
         year = 2023,
        month = apr,
       volume = {520},
       number = {2},
        pages = {2473-2483},
          doi = {10.1093/mnras/stad169},
archivePrefix = {arXiv},
       eprint = {2209.14542},
 primaryClass = {astro-ph.HE},
       adsurl = {https://ui.adsabs.harvard.edu/abs/2023MNRAS.520.2473M}
}

@ARTICLE{Liebendoerfer01,
       author = {{Liebend{\"o}rfer}, Matthias and {Mezzacappa}, Anthony and {Thielemann}, Friedrich-Karl and {Messer}, O.~E. and {Hix}, W. Raphael and {Bruenn}, Stephen W.},
        title = "{Probing the gravitational well: No supernova explosion in spherical symmetry with general relativistic Boltzmann neutrino transport}",
      journal = {\prd},
         year = 2001,
        month = may,
       volume = {63},
       number = {10},
          eid = {103004},
        pages = {103004},
          doi = {10.1103/PhysRevD.63.103004},
archivePrefix = {arXiv},
       eprint = {astro-ph/0006418},
 primaryClass = {astro-ph},
       adsurl = {https://ui.adsabs.harvard.edu/abs/2001PhRvD..63j3004L}
}

@ARTICLE{Kuroda22,
       author = {{Kuroda}, Takami and {Fischer}, Tobias and {Takiwaki}, Tomoya and {Kotake}, Kei},
        title = "{Core-collapse Supernova Simulations and the Formation of Neutron Stars, Hybrid Stars, and Black Holes}",
      journal = {\apj},
         year = 2022,
        month = jan,
       volume = {924},
       number = {1},
          eid = {38},
        pages = {38},
          doi = {10.3847/1538-4357/ac31a8},
archivePrefix = {arXiv},
       eprint = {2109.01508},
 primaryClass = {astro-ph.HE},
       adsurl = {https://ui.adsabs.harvard.edu/abs/2022ApJ...924...38K}
}

@ARTICLE{Burrows23Black,
       author = {{Burrows}, Adam and {Vartanyan}, David and {Wang}, Tianshu},
        title = "{Black Hole Formation Accompanied by the Supernova Explosion of a 40 M $_{{\ensuremath{\odot}}}$ Progenitor Star}",
      journal = {\apj},
         year = 2023,
        month = nov,
       volume = {957},
       number = {2},
          eid = {68},
        pages = {68},
          doi = {10.3847/1538-4357/acfc1c},
archivePrefix = {arXiv},
       eprint = {2308.05798},
 primaryClass = {astro-ph.SR},
       adsurl = {https://ui.adsabs.harvard.edu/abs/2023ApJ...957...68B}
}

@ARTICLE{Kotake18,
       author = {{Kotake}, Kei and {Takiwaki}, Tomoya and {Fischer}, Tobias and {Nakamura}, Ko and {Mart{\'\i}nez-Pinedo}, Gabriel},
        title = "{Impact of Neutrino Opacities on Core-collapse Supernova Simulations}",
      journal = {\apj},
         year = 2018,
        month = feb,
       volume = {853},
       number = {2},
          eid = {170},
        pages = {170},
          doi = {10.3847/1538-4357/aaa716},
archivePrefix = {arXiv},
       eprint = {1801.02703},
 primaryClass = {astro-ph.HE},
       adsurl = {https://ui.adsabs.harvard.edu/abs/2018ApJ...853..170K}
}

@ARTICLE{blondin03stability,
       author = {{Blondin}, John M. and {Mezzacappa}, Anthony and {DeMarino}, Christine},
        title = "{Stability of Standing Accretion Shocks, with an Eye toward Core-Collapse Supernovae}",
      journal = {\apj},
         year = 2003,
        month = feb,
       volume = {584},
       number = {2},
        pages = {971-980},
          doi = {10.1086/345812},
archivePrefix = {arXiv},
       eprint = {astro-ph/0210634},
 primaryClass = {astro-ph},
       adsurl = {https://ui.adsabs.harvard.edu/abs/2003ApJ...584..971B}
}

@ARTICLE{foglizzo06neutrino,
       author = {{Foglizzo}, T. and {Scheck}, L. and {Janka}, H. -Th.},
        title = "{Neutrino-driven Convection versus Advection in Core-Collapse Supernovae}",
      journal = {\apj},
         year = 2006,
        month = dec,
       volume = {652},
       number = {2},
        pages = {1436-1450},
          doi = {10.1086/508443},
archivePrefix = {arXiv},
       eprint = {astro-ph/0507636},
 primaryClass = {astro-ph},
       adsurl = {https://ui.adsabs.harvard.edu/abs/2006ApJ...652.1436F}
}

@ARTICLE{Couch15Three,
       author = {{Couch}, Sean M. and {Chatzopoulos}, Emmanouil and {Arnett}, W. David and {Timmes}, F.~X.},
        title = "{The Three-dimensional Evolution to Core Collapse of a Massive Star}",
      journal = {\apjl},
         year = 2015,
        month = jul,
       volume = {808},
       number = {1},
          eid = {L21},
        pages = {L21},
          doi = {10.1088/2041-8205/808/1/L21},
archivePrefix = {arXiv},
       eprint = {1503.02199},
 primaryClass = {astro-ph.HE},
       adsurl = {https://ui.adsabs.harvard.edu/abs/2015ApJ...808L..21C}
}

@article{dragomiretskiy2014vmd,
    author={Dragomiretskiy, Konstantin and Zosso, Dominique},
    journal={IEEE Transactions on Signal Processing}, 
    title={Variational Mode Decomposition}, 
    year={2014},
    volume={62},
    number={3},
    pages={531-544},
    doi={10.1109/TSP.2013.2288675}
}
\end{document}